\documentclass{llncs}
\usepackage{amsmath}
\usepackage{amssymb}
\usepackage{graphicx}
\usepackage{varioref}
\usepackage[a4paper, total={6in, 8in}]{geometry}
\usepackage{placeins}
\usepackage{hyperref}
\usepackage{newpxtext}
\usepackage{tikz}
\usepackage{xspace}

\title{The Chonkers Algorithm:\\Content-Defined Chunking with Provable Strict Guarantees on Size and Locality}
\author{Benjamin Berger \orcidID{0000-0002-4028-0161}}

\newcommand{\abs}[1]{\left\vert{#1}\right\vert}
\makeatletter
\newcommand{\ie}{{i.e.}\@ifnextchar.{\!\@gobble}{}}
\newcommand{\eg}{{e.g.}\@ifnextchar.{\!\@gobble}{}}
\makeatother
\newcommand{\skipAll}[1]{}
\newcommand{\cc}[1]{\multicolumn{1}{|c|}{{#1}}}
\newcommand{\ccnf}[1]{\multicolumn{1}{c}{{#1}}}

\newcommand{\coolS}{
\begin{tikzpicture}[scale=0.06,baseline=1px]
  \draw (1,0) -- (0,1);
  \draw (1,0) -- (2,1);

  \draw (0,1) -- (0,2);
  \draw (1,1) -- (1,2);
  \draw (2,1) -- (2,2);

  \draw (0,3) -- (1,2);
  \draw (1,3) -- (2,2);

  \draw (0,2) -- (0.5,2.5);
  \draw (2,3) -- (1.5,2.5);

  \draw (0,3) -- (0,4);
  \draw (1,3) -- (1,4);
  \draw (2,3) -- (2,4);

  \draw (0,4) -- (1,5);
  \draw (2,4) -- (1,5);
\end{tikzpicture}
}
\newcommand{\au}{\coolS}

\newcommand{\We}{I\xspace}
\newcommand{\we}{I\xspace}

\begin{document}

\maketitle

\begin{abstract}
This paper presents the Chonkers algorithm, a novel content-defined chunking method providing simultaneous provable strict guarantees on chunk size and edit locality. Unlike existing algorithms such as Rabin fingerprinting and anchor-based methods, Chonkers achieves bounded propagation of edits and precise control over chunk sizes. \We describe the algorithm's layered structure that allows for combination with other chunking algorithms, the theoretical guarantees it provides, implementation considerations, and introduce the Yarn datatype, a deduplicated, merge-tree-based string representation benefiting from Chonkers' strict guarantees. Finally, \we experimentally compare Chonkers' ability to deduplicate versioned data to other algorithms and evaluate Chonkers on three corpora with respect to the actually occurring chunk sizes and edit locality, and find that it performs much better in practice than the proved guarantees.
\end{abstract} 


\section{Introduction}

Content-defined chunking (CDC) partitions an input data stream into variable-size segments, called \emph{chunks}, whose boundaries are determined solely by the data content. CDC underpins many systems that need to detect, store, or transmit only changed regions of large datasets, including:
\begin{itemize}
  \item Backup and archival systems (e.g., LBFS \cite{zhu08}, Venti \cite{quinlan02})
  \item Distributed and deduplicated file systems (e.g., Tahoe-LAFS \cite{wilcox08})
  \item Substring search and delta encoding frameworks \cite{broder93}
\end{itemize}

Two fundamental properties determine CDC effectiveness:
\begin{description}
  \item[Size guarantees:] Chunks should lie within specified size bounds to avoid excessive metadata overhead (too small) and to limit reconstruction cost (too large).
  \item[Locality guarantees:] A local edit (insertion, deletion, modification) should only affect chunk boundaries within a bounded neighborhood, ensuring minimal effect propagation.
\end{description}

Most existing CDC methods optimize for one guarantee at the expense of the other:
\begin{itemize}
  \item \textbf{Rabin fingerprinting} ensures that identical content always produces identical chunk boundaries (strict locality), but only \emph{expected-case} size bounds, which adversarial inputs can violate \cite{rabin81,broder93}.
  \item \textbf{Anchor-based methods} (e.g., Gear \cite{xia16}, FastCDC \cite{xia20}) achieve high throughput and average-case size targets, but cannot prevent boundary shifts arbitrarily far after local edits (no strict locality) \cite{gregoriadis24}.
\end{itemize}

In this paper, \we introduce \emph{Chonkers}, a content-defined chunking algorithm that simultaneously enforces:
\begin{description}
  \item[Strict size bounds:] Loosely speaking, each chunk's weight lies within known fractions of a tunable \emph{absolute unit}, regardless of input.
  Two caveats apply:
  \begin{itemize}
  \item Periodic chunks are allowed to be heavier, but this does not hurt performance because their periodicity can be exploited to store and process them efficiently. The period will be less than one absolute unit.
  \item Isolated chunks surrounded by such overly heavy chunks can have arbitrarily low weight, but this likewise does not hurt performance because the surrounding heavy periodic chunks more than compensate for it in their effect on average chunk weight and chunk count.
  \end{itemize}
  \item[Strict locality bounds:] any single edit can perturb at most a constant number of neighboring chunks, regardless of how adversarial the input is. Strictly speaking, the number is not constant, but involves the iterated logarithm of the absolute unit. But for all realistic applications the absolute unit is less than $2^{2^{126}}$. With this bound, the contribution of the iterated logarithm is bounded by $7$, so this is only theoretical.
\end{description}
Chonkers is built as a hierarchical, layered algorithm with deterministic, priority-based merging of adjacent chunks. \We also present \emph{Yarn}, a deduplicated merge-tree data structure leveraging Chonkers for efficient storage, comparison, and substring operations.

\section{Background and Related Work}

\subsection{Rabin-Fingerprint-Based CDC}
Rabin-fingerprint CDC \cite{rabin81} uses polynomial hash checks 
to define chunk boundaries deterministically. 
It satisfies \emph{strict locality}: 
identical content far from an edit is chunked identically. 
However, its chunk sizes are only \emph{expected} to meet target bounds; 
adversarial inputs can produce extremely small or large chunks \cite{broder93,rabin81}. 

Systems like LBFS \cite{zhu08,tridgell96} rely on this approach for deduplication 
and synchronization.

\subsection{Anchor-Based CDC}
Anchor-driven chunkers such as Gear \cite{xia16} and FastCDC \cite{xia20} select boundaries based on rolling-hash anchors to approximate target sizes while improving throughput by factors of 3–10 over Rabin-based methods. Despite their speed, they lack any formal locality guarantees: a single-byte change can shift all subsequent boundaries. Recent benchmarks confirm the absence of stability guarantees in anchor-based CDC \cite{gregoriadis24}.

\subsection{Hybrid and Two-Threshold Methods}
Multi-stage and two-threshold chunking algorithms (e.g., TTTD \cite{jasim18}, 
Anchor-Driven Subchunk \cite{romanski17}) combine coarse and fine-grained passes 
to reduce variability in chunk size. 
The Asymmetric Extremum algorithm \cite{zhang16} further smooths size distributions. However, none provide proofs of both strict size and locality bounds, leaving systems vulnerable to adversarial inputs.

Chonkers addresses these shortcomings by deterministically bounding both guarantees and structuring chunking in composable layers. The subsequent sections detail the algorithm, theoretical analysis, and practical considerations.

\section{Definitions and Terminology}
In the interest of a consistently themed terminology, \we propose following definitions:

In the context of the Chonkers algorithm, the number of bits in a chunk $c$ is called its \emph{weight} and written ``$\abs c$ bits''. 
This leaves the terms \emph{length} or \emph{size} to be used by applications, for example to measure byte or character counts.
Without the unit specification ``bit'', $\abs c$ is measured in absolute units.

The \emph{absolute unit} $\au$ is a parameter limiting how large the chunks can be that Chonkers attempts to form normally.
We usually specify chunk weights relative to the absolute unit, and do not mention the unit. 

Chonkers forms chunks by merging smaller chunks, starting from an initial \emph{proto-chunking}. Proto-chunks can for example be bytes, UTF-8 encoded characters, DNA base pairs, LLM tokens or the output of another chunking algorithm.

Chunks are classified according to their weight:
\begin{itemize}
\item \emph{Megachonkers} are chunks with weight $\ge 1$. 
\item \emph{Heftychonks} are chunks of weight $\ge \frac12$ but $< 1$.
\item \emph{Fine bois} are chunks of weight $\ge \frac14$ but $< \frac12$.
\item \emph{Kittens} are chunks of weight $< \frac14$.
\end{itemize}

Known exact periodic repetitions of $2$ or more copies of the same bit content are termed \emph{caterpillars}. The repeating bit pattern weighing one period is called the caterpillar's \emph{segment}. Periodic parts of the input are not guaranteed to result in a caterpillar, and if they do, the caterpillar might not span the maximum number of periods. The term applies only to periodic parts that have explicitly been identified as such.

A megachonker can only exist as a caterpillar of period $< 1$, or if it was already present in the proto-chunking.
Nevertheless, we will also use the term to talk about hypothetical chunks that will not be formed by Chonkers, precisely because they would be too heavy.

Two adjacent chunks $l$, $r$ that can be merged without resulting in a megachonker, i.e., $\abs l + \abs r < 1$, are said to be \emph{heckin' with each other}, the boundary between them is \emph{heck'd} by them, and a chunk that is heckin' with at least one neighbor is called a \emph{heckin' chonker}.

The bits that make up a chunk are called its raw bit content.
Each chunk also has an augmented bit content, which is formed from the raw bit content by prepending a binary representation of the chunk's weight, and optionally a hash code.
Bits are considered ordered left to right when it comes to the data, and ordered least significant to most significant when it comes to binary representation of numbers. Keep in mind that this contravenes the usual big-endian convention of writing numbers in a place-value system.

When talking about the neighbors of a chunk, we say ``all neighbors'' instead of ``both/two neighbors'' to account for the fact that a chunk can be at the beginning and/or the end of the sequence, where it will have less than two neighbors.

The \emph{diffbit} of two bit sequences is a number $2\cdot i + d$, where $i$ is the zero-based index of the first bit position where they differ and $d \in \{0, 1\}$ indicates the direction of the change: If the bit at index $i$ is $0$ in the left sequence, but $1$ in the right, then $d=1$, whereas if the bit changes from $1$ to $0$ between the left and right sequence, then $d=0$.

\section{The Chonkers Algorithm}
Chonkers processes data in composable layers, each having at most double the absolute unit weight from the previous. 
That is, if $\au_i$ denotes the absolute unit that applies to layer $i$, then $\au_{i} < \au_{i+1} \le 2\cdot \au_{i}$
Exactly doubling the absolute unit between layers is optimal while still enabling the provable weight guarantees, but applications might opt for a lesser factor in some layers in order to achieve a desired value for the absolute unit of the final layer. More than doubling the absolute unit may be an option if weight guarantees are not deemed important.

Each layer contains three sequential phases:
\begin{itemize}
\item The balancing phase
\item The caterpillar phase
\item The diffbit phase
\end{itemize}
The inputs and outputs of successive phases form progressively coarser chunkings of the data.

The output of each phase is the input to the next phase, or to the first phase of the next layer.

This layered design guarantees that the postconditions of one layer meet the preconditions of the next layer, enabling an inductive proof of both weight and locality bounds.

\subsection{Algorithm Phases}

\subsubsection{Balancing Phase:} 
In the balancing phase, an attempt is made to merge chunks with at least one of their neighbors if they are lighter than all their neighbors. 
The balancing phase is needed to prove the weight guarantees, but can also be utilized to more aggressively merge chunks\footnote{See subsection \vref{moreBalance}}. This will improve the locality properties of Chonkers to be in many cases better than the provable worst case.

A chunk that is lighter than its neighbors will preferentially be merged with its right neighbor, at priority $0$. An attempt will also be made to merge it with its left neighbor, at priority $1$. Since the priority-based merging mechanism will also be used in the third phase, we will describe it later for both phases, in subsection \vref{priomerge}. This scheme to assign merge priorities is well-defined because there can be no two adjacent chunks so that each is lighter than its neighbors, so each boundary that is assigned a merge priority is the boundary of only one locally minimal chunk.

For the balancing phase, the \textbf{precondition} which we need to prove the weight guarantees is: 
There are no two consecutive kittens.

If this precondition is met, then after the balancing phase, the following \textbf{postcondition} will hold: There is no kitten that is also a heckin' chonker.

\subsubsection{Caterpillar Phase:} 
The caterpillar phase scans for any number of consecutive chunks with equal bit content, and merges as many as possible of them into a caterpillar.

Optionally, we can exclude forming caterpillars of megachonkers. Such megachonkers can only be overweight proto-chunks\footnote{They cannot be caterpillars formed by Chonkers itself in earlier layers, because if they were equal, they would have been a single caterpillar, not several.}.
But the only benefit of this is to make the maximum weight guarantees easier to state.

The caterpillar phase preserves the weight related postcondition of the balancing phase, and provides the following additional \textbf{postcondition}: There are no two consecutive chunks with exactly equal bit content that are heckin' with each other\footnote{This last stipulation is unnecessary if we allow forming caterpillars from overweight proto-chunks. In that case, there will be no consecutive equal chunks at all.}.

This postcondition is necessary for the diffbit phase to work, as diffbits are only defined for unequal bit sequences.

\subsubsection{Diffbit Phase:} 
In the diffbit phase, Chonkers computes the diffbits of the augmented chunk contents of all pairs of chunks that are heckin' with each other. Remember that this means we can merge them without forming a megachonker, because their combined weight is less than one absolute unit.

These diffbits are well-defined: Thanks to the caterpillar phase, the two consecutive chunks will be different, and if they differ in weight, then the first index where they differ will occur no later than in the binary representation of the weight that is prefixed to the raw bit content, preventing the possibility of comparing bits at indices that don't exist in both chunks.

We ascribe the diffbit to the left chunk of the pair, and call it the first order diffbit of that chunk.

Chunks that are not heckin' with their right neighbor also get annotated with diffbits, but in the interest of limiting the spreading of causal influences, we do not want these to actually depend on the right neighbor. Hence, we use the diffbit that we would obtain for a fictitious right neighbor that differs at bit index 0. That is, the diffbit is $0$ if the chunk's zeroth bit is $1$, and otherwise the diffbit is $1$.

This guarantees that any two consecutive diffbits are distinct. If two consecutive diffbits were equal, it would mean that over the course of three chunks, the bit at same bit index changes twice in a row in the same direction. But as there are only two possible values for a bit, this is impossible.

The distinctness of consecutive diffbits allows us to iterate this construction: The second order diffbit of a chunk $c$ is the diffbit of the binary representations of the first order diffbit of $c$, and of the first order diffbit of $c$'s right neighbor\footnote{Or, if $c$ is not heckin' with its right neighbor, the appropriate fictitious substitute in the same manner as we did it for the first order diffbits.}.

In the same manner, higher order diffbits are defined recursively: The order $(n+1)$ diffbit of a chunk is the diffbit of its order $n$ diffbit and the order $n$ diffbit of its right neighbor.
How many orders of diffbits do we need? Suppose the augmented bit content is shorter than $2^{127}$ bits.
Then
\begin{itemize}
\item $128$ bits are needed to represent the first order diffbit: $127$ bits for the index where the change occurs, plus one to indicate the direction.
\item $8$ bits are needed to represent the second order diffbit: $7$ bits for the index where the change occurs (in the range from $0$ to $2^7 -1$), plus one to indicate the direction.
\item $4$ bits are needed to represent the third order diffbit: $3$ bits for the index where the change occurs, plus one to indicate the direction.
\item $3$ bits are needed to represent the fourth order diffbit: $2$ bits for the index where the change occurs, plus one to indicate the direction.
\item $3$ bits are needed to represent the fifth order diffbit: $2$ bits for the index where the change occurs, plus one to indicate the direction. However, since the index is at most $2$, only the values $0$ to $5$ will occur.
\end{itemize}
A chunk weight of $2^{127}$ bits appears preposterously large, but the Chonkers algorithm, when used with deduplication and our Yarn datatype, can in principle create even heavier chunks, for example by iteratedly concatenating a string with itself. So, for very heavy absolute units, we might need a sixth order diffbit. A need for a seventh order diffbit can safely be ruled out: The number of operations needed to construct such a situation, and the number of bits needed to represent the chunk weights in bits, would exceed the constraints of physical space-time.

Having now obtained the highest order diffbit for each chunk $c$ where we will need it, we associate it as a merge priority with the common boundary of $c$ and its right neighbor, provided that boundary is heck'd.

We then proceed to merge chunks based on these priorities, as explained in subsection \vref{priomerge}.

The \textbf{preconditions} for the diffbit phase are postconditions of the caterpillar phase. If they are fulfilled, then the diffbit phase has the following \textbf{postcondition}: No two consecutive chunks have weight $< \frac12$ (\ie, no two fine bois or kittens remain adjacent). No two consecutive chunks are fine bois or kittens.

\subsubsection{Computing Diffbits Efficiently}
By structural deduplication we maintain a DAG of chunk derivations, where each node points to its two children -- or more, in the case of a caterpillar. Deduplicating chunks using a hash table as soon as they are formed will yield a directed acyclic graph, with much structure shared between the subgraphs reachable from similar chunks. Conceptually, each chunk is a tree recording the history of how it was merged from smaller chunks, but deduplication makes storage and operations more efficient.

Particularly, it makes determining the first order diffbit efficient: Even if the first difference occurs at a very high index, the locality guarantees that apply at each layer of the tree mean that all but a constant number of nodes near the beginning, and also a bounded number near the searched index, will need to be checked at each tree layer; most will be identical and will not need descending into.

Finding the diffbit of the augmented bit contents is further sped up by prepending a hash code as part of the augmented bit content. Then looking at the actual raw bit content is only needed in the rare case of a hash collision.

Finding the diffbit in the portions of the raw bit content where skippable identical subtrees are not available is more complicated. First, we generalize the problem to finding the diffbit starting from an offset into each bit sequence, and ending with the last common position counted from the offset. With this, a mutual recursion between the two trees is used to descend as far as needed to find the differing bit or to verify that both suffixes starting at the offsets have the same content. As already mentioned, the locality guarantee enables skipping most of the nodes. For the details of the mutual recursion, see the reference implementation.

\subsubsection{Concerning Hashes}
Hash codes are also needed for efficient deduplication. 
To avoid processing all data in a newly formed chunk, it should be possible to compute its hash value from the hash values of its constituents.

There exist hash functions that are homomorphisms from the set of strings to a monoid, so that the hash code of the concatenation of two strings can be computed from the hashes of the parts. One class of such hash functions is based on the special linear group over finite fields \cite{MullanTsaban2016}.

A less collision-resistant, but possibly cheaper alternative is to use a Rabin fingerprint $h$: 
Being a hash function that is a polynomial with a fixed base $b$ and exponents provided by the bits to be hashed, the fingerprint $h(uv)$ is simply $h(u) + b^{\frac{\abs{u}}{\text{bit}}}h(v)$, with operations performed in the appropriate ring.

\subsection{Priority-Based Merging Mechanism}
\label{priomerge}
Priority-based merging of chunks is used by both the balancing phase and the diffbit phase and works the same way in both cases, after the priorities have been assigned.

Priority-based merging iterates through the possible priority values. When processing the priority $p$, it eliminates all chunk boundaries annotated with that priority by merging the adjacent chunks, provided that none of the following rules prevent it:
\begin{itemize}
\item No-megachonkers rule: This merger would not result in a megachonker, that is, only boundaries which are still heck'd will be eliminated.
\item Determinism rule: To ensure well-defined local determinism regardless of the order in which the prospective binary mergers are considered, the merge priority that was ascribed to the right boundary of the right chunk must differ from $p$. 
In other words, if there is a consecutive run of possible mergers of equal priority, only the rightmost merge can be performed.
If the possible mergers are not processed from left to right, then the implementation must make sure that, after the rightmost boundary with merge priority $p$ disappears due to the merger, the boundary to the left of it will not be merged later at the same priority. 
\end{itemize}

The merged chunk inherits the merge priorities of the left boundary of its left constituent, and the right boundary of its right constituent, and can therefore in principle undergo further mergers. The priorities are inherited even in the case that the chunk boundary is no longer heck'd and will therefore not be eliminated (due to the no-megachonkers rule). This is necessary as the merge priority can still be relevant to the determinism rule, and removing it would lead to unwanted spread of causal influences from the merger, weakening the attainable locality guarantees.

\subsection{Proofs}
The proofs of locality and weight guarantees work by induction and can be assembled by plugging the postcondition of each phase into the precondition for the next, proving that the phase's postconditions follow freom the preconditions, and performing induction over the layers while doubling the absolute unit from layer to layer. The induction starts by making the absolute unit for the first layer no more than twice as heavy as the smallest possible proto-chunk. The details of the proofs and the statement of the proven guarantees are found in appendix \vref{sectionProofs}.

\skipAll{
\subsection{Additional Aspects}
Section \vref{sectionTweaks} presents variants and modifications of this core algorithm.
Section \vref{sectionTree} describes how to rebuild the merge tree after a local change by only re-running the parts of the algorithm that can be affected by it.
}

\section{Chonker Tree}
\label{sectionTree}

The Chonkers algorithm can be described in terms of transforming an input list of chunks into an output list of chunks by merging chunks into heavier chunks. 
If these chunks store references to their constituent chunks, then the lists are naturally forests. Iterating the Chonkers layers until only a single chunk remains will give us a tree, with the proto-chunks as its leaves.

Given such a chonker tree, we can iterate through the input and output lists for each layer/phase combination, provided that the chunks also remember at which layer and phase they were formed. This gives us the ability to quickly compute the chonker tree of a modified sequence if we have the tree for the original, by replaying only the parts of the algorithm's execution trace that may have been affected by the change.

The following description of the chonker tree rebuilding algorithm is only a sketch of one way how to do that. We will not go into details; for those, we point the reader to the reference implementation at \cite{Berger2025Chonkers}.

We generalize the operations of concatenation, deletion, insertion, replacement and de-novo-creation on chonker trees\footnote{For relatively short sequences. De novo creation of the trees of long sequences is better served by a streaming and/or parallelized implementation.}. 
The Chonker tree rebuilding algorithm uses a zipper\cite{Huet1997Zipper} into a chonker tree of some string that points to the rightmost node of a desired prefix. Likewise, there is a zipper specifying the desired suffix. 
Lastly, there is a list of proto-chunks that should be inserted in between the prefix and the suffix. 

The algorithm works through the layers and phases from the bottom-up. Initially, the zippers point at leaves, and the list of proto-chunks gets stored as a doubly linked list of items. These items have a reference to the chunk they manage, but also store the merge priority associated with its right boundary and flags indicating whether the chunk is \emph{tainted} by having a possibly wrong structure.

At the beginning of each balancing phase and diffbit phase, the left zipper is moved left through the input list of chunks for that phase, \footnote{Although the zipper points at a node in a tree, within a phase it is better to conceptualize it as pointing at a position in the list represented by the level in the tree corresponding to the current phase's input list of chunks.} and the nodes it no longer points to are added to the left end of the linked list. It is moved far enough that we can be sure (thanks to the locality guarantees) that the structure of the list prefix represented by the zipper will not be affected by the current modification. Likewise, the right zipper is moved right. 

The leftmost and rightmost items in the linked list are marked as tainted. In the case of the diffbit phase, as many rightmost items are tainted as we cannot compute the highest order merge priority for, because they don't have enough right neighbors in the linked list. The taint represents the possibility that a chunk is affected by something outside of the linked list.

Next, we compute the merge priorities based on the chunks in the linked list, accepting that they might be wrong at the ends of the list.
The priority-based merging is applied to the linked list, but the taint must be spread like a causality infection: If a tainted chunk is merged with something, the result will be tainted, too. And if a merger is prevented by some aspect (weight or merge priority) of a tainted chunk that was not already present at the beginning of the phase, then the non-merged chunks also gets tainted.

At the end of the phase, the tainted items are removed from the linked list and the zippers are moved back according to the weight of the tainted items. 

The caterpillar phase tries to form caterpillars from the chunks in the linked list, but then also accesses the left and right zippers at the 
level of the phase's output to see if they point into a caterpillar that can be unified with a caterpillar with the same segment, or a single chunk equal to the segment, in the linked list. If the linked list is empty, the possibility that the left and right zippers both point to caterpillars of the same segment, or to single chunks equal to that segment, and consequently can be merged into a single caterpillar, must also be considered.

When using the Chonkers algorithm to just one-off chunk a sequence, the chunk deduplication can be based on the raw bit content alone. But if the intention is to later apply the tree rebuilding algorithm to it, the hash function and equality predicate used for the deduplication must also account for the tree structure and the layer and phase at which chunks were formed. They need not descend into the tree structure, because the constituents of a merged chunk can be assumed to be deduplicated. Hence, the hash function and equality predicate can operate on memory addresses\footnote{Implementations in higher level languages that do not expose memory addresses can use reference equality and identity hash codes.}. 

Caterpillars cannot simply be represented by a prototypical chunk for the segment together with the period count. It is necessary for the tree rebuilding algorithm to have access to the full list of different tree structures of the constituent nodes. However, thanks to the locality guarantees, these tree structures will only differ near the head end and the tail end of the caterpillar. Therefore, caterpillars can be represented and processed efficiently using run-length encoding. \cite{shirani2008data}

\section{Tweaks and Variations}
\label{sectionTweaks}
This section discusses small improvements and variations that do not affect the core of the algorithm. Two of these have been implemented in the reference implementation and used for the experiments. They are described in subsections \vref{moreBalance} and \vref{twoNodeTypes}. The others can be seen as suggestions for further development.

\subsection{More Aggressive Merging in the Balancing Phase}
\label{moreBalance}
The balancing phase was introduced to obtain provable weight guarantees. 
But it is also is an opportunity to perform more mergers with a lower radius of causal influence than the diffbit phase.
If we can merge more chunks in the balancing phase, there will be fewer heck'd boundaries in the diffbit phase.
This is good, because non-heck'd boundaries block the causal influence from both cascades and diffbit computation. 

There are two ways to merge more chunks in the average case that do not affect proofs for the guarantees:

First, if we find that two chunks have equal weight, we can compare them lexicographically. A chunk is then also considered smaller than its neighbor if it is of equal weight, but lexicographically smaller. The chunks that will be considered for merging with their neighbors will still include the locally lightest. For the proof of the weight guarantees, we are actually just interested in kittens. But merging heavier chunks that are lighter than their neighbors is fine too. Considering even more chunks by comparing lexicographically also does not interfere with merging kittens that are heckin' chonkers, because these additional chunks will not be neighbors of kittens, which are strictly lighter than their neighbors according to the precondition. By including a sufficiently pseudo-random hash function in the comparison,\footnote{for example, by reusing the diffbit phase's code for computing the diffbit of bit contents augmented with a hash function} in the expected case about every third chunk will be considered smaller than both its neighbors \cite{romik2011local}.

Second, we can relax the comparison with either the left or the right neighbor to be non-strict. This way, out of an extended local minimum, \ie, of a run of equally heavy chunks that are lighter than the chunks neighboring the run, we get to merge two. Note that this tweak does not really make sense if we also compare chunks lexicographically as described in the previous paragraph: Then such an extended local minimum would be made up of equal chunks, and we are better served by the caterpillar phase dealing with it.

\subsection{Streaming}
When chunking a long file with Chonkers, it is disadvantageous to first wrap each byte into a proto-chunk, and then to merge chunks layer by layer, phase by phase and priority by priority. A conceivably better way in terms of memory access patterns is to scan the file left to right, and passing streams of chunks through a stack of stream processors. These stream processors each just buffer enough lookahead chunks to do their job: 
\begin{itemize}
\item The processor for computing diffbit merge priorities will need to look ahead up to seven chunks, and will emit chunks annotated with the merge priority of their right boundary. 
\item Subsequent priority-based-merging processors for each priority can then buffer just two chunks and base the decision whether to merge them on their weights and merge priorities. 
\item The processor for the balancing phase merge priorities needs to buffer three chunks, in order to check that the middle one is smaller. It also annotates the chunks with merge priorities before passing them to subsequent merging processors.
\item A processor for the caterpillar phase will just buffer two chunks, the left of which might be a caterpillar that the right one gets appended to as long as it is equal to the caterpillar's segment. 
\end{itemize} 
When building a chonker tree, processors for more layers can be added as needed to the top of the stack as the depth of the tree grows.
\subsection{Mirror Version}
So far, the algorithm has been described with a left-leaning bias of the causal influences: The diffbits depend on the right neighbors of a chunk, and the determinism rule checks the merge priority of the boundary to the right of the current boundary. 
By basing the diffbits on the left neighbors instead, or by having the determinism rule check the left neighbor, and ascribing the merge priorities to the left boundaries of chunks, we get a mirror version of Chonkers with a right-leaning bias. This might be slightly preferable for the streaming scheme described in the previous subsection, as it reduces latency and buffer size. Especially the buffer for computing diffbit phase merge priorities now only needs to contain two chunks, instead of seven\footnote{Or $O(\log^* \au)$, theoretically}.
\subsection{Parallelization}
The locality guarantee enables easy parallelization: A long file can be divided into sections, and these sections extended to all sides so that they overlap. The extended sections are then chunked separately in parallel. The lengths of the left and right extensions are calculated so that, for the target maximum chunk weight, the causal influences on chunk boundaries from the end of the extended section cannot reach beyond the original section boundary. The chunking results can thus be merged by essentially taking the union of the set of chunk boundary positions that lie within the original section. Caterpillars with equal segments on both sides of the section boundary may need to be fused.
\subsection{Combining with an Efficient Proto-Chunker}
The pure Chonkers algorithm with hashcode-augmented bit contents uses a constant number of operations per input proto-chunk, unless hash collisions occur. Still, it is probably not competitive in terms of raw throughput when compared to existing chunking algorithms. Its unique feature is that it gives both strong weight guarantees and strong locality guarantees. 

Due to Chonkers' layered architecture, it is easy to combine with another chunking algorithm that is more efficient. If that other chunking algorithm offers strong locality guarantees, but only probabilistic weight guarantees, the combined algorithm can be made to offer both guarantees while being efficient in the average case.

This works as follows: The efficient but probabilistic algorithm is given a target weight so that most chunks it will produce are close to, but lighter than, the ultimate absolute unit that Chonkers will strive for. If the efficient algorithm produces chunks shorter than the ultimate absolute unit, they will be handed as proto-chunks to Chonkers. In most cases they will weigh a sizable fraction of the absolute unit and so Chonkers has little to do. If the efficient algorithm produces, or is about to produce\footnote{We will notice that as soon as it has not produced a chunk boundary for long enough}, a megachonker, we have a few options:
\begin{itemize}
\item Backtrack and run another efficient algorithm with strong locality guarantees, or the same algorithm with different parameters, on the substring covered by the overweight chunk. 
\item Backtrack and run an algorithm for detecting periodicity on the overly long chunk. Convert periodic sections into proto-chunks that are caterpillars. Retry chunking any non-periodic sections. This addresses a common failure case for efficient hash-based chunkers, in which data is periodic. When the periodic section is converted into a caterpillar, care must be taken that its segment is independent of the shift at which the periodicity started. This may require excluding a prefix and/or suffix of the periodic section from the caterpillar. For example, one could choose the period to start at the shift that makes it lexicographically smallest.
\item If all else fails, we have to hand the individual bytes\footnote{or whatever units we naturally can divide the file into} as proto-chunks to Chonkers until the more efficient algorithm finds a boundary again. This will negate the efficiency gain we hoped for by using the efficient proto-chunking algorithm, but will be rare in the expected case.
\end{itemize}
\subsection{Separation of Structure and Content}
\label{twoNodeTypes}
Most values computed from chunks, such as the weight, diffbits, hash values, and the raw bit content, as well as generally any custom monoid values derived from the underlying bit sequence via a homomorphism, do not depend on the structure of the merge tree that led to the chunk.
The structure is only needed to rebuild the chonker tree. 
It seems therefore a good idea to separate these two concerns by using structure nodes and content nodes, and deduplicate them separately using the appropriate hash function and equality predicate. 

Then all structure nodes with the same content can share a reference to a content node. The content node in turn will have a reference to any one of the structure nodes that rely on it, so that it can access the raw bit content for computations of diffbits, or access constituent nodes for computing monoidal values. Since monoids are associative, it does not matter which way the structure node splits the content.

One caveat here is that the prototypical structure node referred to by a content node might not be in use anywhere else, wasting memory. The memory wasted can be more than what is needed for one structure node, because its subtrees and/or their content nodes might also not be used anywhere else. A simple workaround for this is to change the content node's reference pointing to its prototypical structure node each time we look up the content node of a structure node $k$. Changing the reference to point to $k$ has the effect that it now points to a node that is known to be in use.
Care must be taken when doing a computation that depends on this reference, as it may be changed concurrently at any time, albeit to a structure node of equal content. Still, any method that asks a content node for its structure node prototype should better do so only once.

\subsection{Compact Storage}
A chonker tree that has a leaf for each character will cause a lot of overhead when the nodes are represented using dynamically allocated memory. And if chonker trees were to be used to implement a file system, it would be a wasteful use of the usual block structure of storage devices if every node got its own block.

Hence, it would be advantageous to compact nodes. A compacted node represents a part of a chonker tree using fewer separate objects. It consists of an array containing either the serialized bit representation of its content, or references to the roots of subtrees\footnote{With the exception of subtrees that are repeated with the exact same structure inside a many-segmented caterpillar; these may be stored only once, but require special annotation that the stored sequence of bits or references cannot be used as-is. Heavy caterpillars cannot occur too often within a compacted node, as they will not undergo mergers until the absolute unit catches up with their weight. Thus a compacted node containing a heavy caterpillar will be made up mostly of that caterpillar.}. Additionally, it has a compact index that enables reconstructing the structure of the tree part that was collapsed into the compacted node.
This index will not be needed for all operations, but if it is needed, for example during tree rebuilds, the original structure can be restored. The compacted node may also keep a memory-sensitive reference to the ordinary, non-compact version of the tree. Thus nodes that were recently decompacted will not need to be decompacted again, provided memory is available to store the decompacted tree. This memory can be freed at any time, should the need arise, by some memory management system such as a garbage collector.

Here is a suggestion for one way to structure the compact index: The tree part that gets compacted extends vertically over one or more whole layers, and has a single top node. The indices for layers are encoded bottom up. The index for a layer contains an entry for each chunk boundary in the part of its input covered by the tree. Entries consist of 4 bits (one nibble) encoding 
\begin{itemize}
\item Whether the chunks were not merged in this layer (1 code needed)
\item Merge priority for the balancing phase. (2 codes needed)
\item Merge priority for the diffbit phase. (6 codes needed)
\item Format info for caterpillars (7 codes available)
\end{itemize}

Caterpillars are encoded in multiple bytes. For longer caterpillars, the nodes in the stored layer input may be only the segments that make up the structurally different runs in the run length encoding of the caterpillar. Depending on whether this is done, as well as on the number of segments and the number of runs of structurally equal segments, different formats can be chosen to specify the number of runs and the run lengths, disambiguated by the code used in the initial nibble. Deciding on the details of caterpillar encoding will require further work of analyzing the distribution of caterpillar structures in typical data. 

The index will typically have uneven distribution of nibbles, as some merge priorities are more likely than others, and about every second entry will indicate that the chunks were not merged, as the maximum and average weight of chunks roughly doubles per layer. Hence, the index might benefit from a simple fixed entropy coding compression, such as a Huffman code table depending on the previous two nibbles.

\section{The Yarn Datatype}
Yarn is a datatype for character sequences that encapsulates a chonker tree with characters at its leaves.
Like rope datatypes\cite{Boehm1995Ropes}, it represents the data as a balanced tree, enabling $O(\log^* n \cdot \log n)$ expected time complexity for all basic operations.
Unlike other rope datatypes, the tree structure is unique and well-defined for each sequence. It does not depend on the history of how the Yarn was assembled using these operations. This makes equality comparison trivial. Efficient lexicographic comparison can be realized by reusing the algorithm for computing the diffbit of raw bit contents. This will not only determine which of two Yarns is lexicographically smaller, but also provide the index of the first character where they differ.
\subsection{Example for Why Yarns are Different}
Consider, for example, the $80$th Fibonacci word\cite{berthe2010combinatorics}. It is aperiodic and over 61 quadrillion characters long. It can be constructed by concatenating Yarns in about a quarter of a second using the unoptimized, single-threaded reference implementation written in Java. Other rope datatypes will usually be faster at this because they use a laxer balancing scheme. 

Fibonacci words have the property that removing the last two characters yields a palindrome\cite{berthe2010combinatorics}. We can verify this by removing the last two characters, reversing the result and comparing the reversed and unreversed Yarns. A Yarn can be reversed by computing, for each node in its chonker tree, the reverse of the represented string from the reverses of its constituents, according to $R(uv) = R(v) R(u)$. Each concatenation requires running the chonker tree rebuilding algorithm, but the results can be cached in the content nodes, and for the $80$th Fibonacci word, less than 260 content nodes are used to represent the palindrome. Hence, the reversal takes only slightly longer than the initial assembly of the Fibonacci word. The comparison is then instantaneous.

Other rope datatypes can also quickly be reversed using a cached monoid,\footnote{provided they are sufficiently deduplicated, which they will be after constructing a Fibonacci word} but the rope resulting from reversing a palindrome will most likely not have the same tree structure, making comparison harder. 

There are datatypes for representing strings based on generative grammars\cite{lohrey2012algorithmics}. These enable efficient palindrome tests, but are not that flexible. They work best for very regular strings that can be described by grammar production rules, and inserting arbitrary text into such a string usually requires a complete rewrite of the grammar. In contrast, Yarn would handle insertion of, \eg, a short palindromic text into the middle of any $n\approx 61$ quadrillion character palindrome as easily as any other operation, requiring only $O(\log^* n \cdot \log n)$ changes to the tree structure. The reversal would then take almost the same amount as without the inserted text, and the equality test is still trivial.

\section{Reference implementation}
The reference implementation \cite{Berger2025Chonkers} comprises bottom-up tree construction, the tree rebuilding algorithm, and the Yarn datatype. It is intentionally not optimized for maximum performance. Streamed or parallel chunking, compact subtree storage, as well as combining with an efficient proto-chunker, are not implemented.

\section{Experimental Results}
\label{xp}
The theoretically provable bounds are one thing, but the behavior of the algorithm in practice is much better.
In this section I evaluate the distribution of chunk weights, the locality of edits on chunk boundaries, and the distribution of phases and merge priorities at which chunks were merged.
For the evaluation, I use three different text corpora:
\begin{itemize}
\item Random strings: $10\,000$ strings of length $10\,000$, consisting of uniformly distributed characters with code points between $0$ and $255$.
\item Linux kernel, commit hash \verb'b19a97d57c15643494ac8bfaaa35e3ee472d41da' \cite{kernel}: $61\,834$ files with extension \verb'.c' or \verb'.h', containing $1\,360\,700\,871$ characters in total. 
\item German fiction \cite{Fischer2017}: $3\,219$ files of plain text, with a total of $1\,075\,375\,991$ characters.
\end{itemize}
\subsection{Algorithm Parameters}
The variant of the chonkers algorithm used in this experiment takes each character represented as a $32$-bit proto-chunk as input. 
From layer $3$ onwards, the augmented bit content is prefixed with a $32$-bit polynomial hash.
The absolute unit for layer $n$ weighs $1+32\cdot 2^n$ bits. 
For each layer, the $5$th order diffbits were used as the merge priorities in the diffbit phase.
The balancing phase optimization\footnote{See subsection \vref{moreBalance}} that compares the augmented bit content of equal weight chunks lexicographically is used.
\subsection{Weight Distribution}
For each layer and string in the corpus, we record the average chunk weight, the standard deviation of chunk weights, the maximum segment weight, and the minimum adjacent pair weight. The segment weight for a caterpillar is its period, and for a non caterpillar it is just its weight. The weight guarantees imply that the segment weight cannot exceed $1$. The weight guarantees imply that the combined weight of two consecutive chunks cannot be less than $1$.

Then the average, standard deviation, minimum and maximum of these values are computed across all strings in the corpus.
The aggregation is not weighted by length, and considers only the layers that are actually present in the chonker tree for a string.
The results are reported in absolute units.

Table \vref{fictionsize} shows the results for the German fiction corpus.
Table \vref{randomsize} shows the results for the random corpus.
Table \vref{kernelsize} shows the results for the kernel source code corpus.

We see that the average weight is quite reliably around $0.7$. Especially in the kernel corpus, the average weight in early layers can exceed $1$, but this is due to the frequent occurrence of caterpillars\footnote{Resulting from indentation, for example}. We can see this because the maximum segment weight never exceeds $1$.

\subsection{Locality Distribution}
For each string in the corpus, nine evenly spaced edit locations are chosen. For each edit location, one character is deleted and then for each layer, the number of characters between the edit location and the nearest position beyond which chunk boundaries at that layer are equal is determined, both to the left and to the right of the edit location.

Then the average, standard deviation and maximum of these values are computed across all strings and edit location choices.
The aggregation is not weighted by length, and considers only the layers that are actually present in the chonker tree for a string.
The results are reported in absolute units.

Table \vref{fictionsize} shows the results for the German fiction corpus.
Table \vref{randomlocality} shows the results for the random corpus.
Table \vref{kernelsize} shows the results for the kernel source code corpus.

We see that the locality is actually much better than the proven bound: The average is always well below $1$, and the maximum does not exceed $5$.

\subsection{Phase Census}
For each string in the corpus, the chonker tree is inspected with regard to the phase and merge priority at which its nodes were created.
The counts are aggregated per layer. Subtrees occurring multiple times are counted individually.

The results are reported as percentages, first aggregated over all layers and then individually for each layer.

Table \vref{fictionsize} shows the results for the German fiction corpus.
Table \vref{randomcensus} shows the results for the random corpus.
Table \vref{kernelsize} shows the results for the kernel source code corpus.

We see that most mergers happen in the balancing phase at priority $0$, followed by mergers in the diffbit phase at priority $0$.
Interestingly, at later layers, the share of balancing phase merge priority $1$ grows considerably.
 
\subsection{Deduplication Benchmark}
To evaluate how Chonkers deduplicates data, the three corpora used so far are not a good choice on account of containing not many long duplicate portions. A better choice here is to use different versions of the same document set. This has the additional benefit of testing the use case of versioned file storage.

\We use ten snapshots of the Linux kernel, from version 6.0 to version 6.9, as the dataset\footnote{Files smaller than $16$~kiB were removed, as they don't really involve much chunking and would distort the chunk size statistics.}, and use DedupBench \cite{liu2023dedupbench,github_dedup-bench_2025} to compare Chonkers against the other algorithms implmenented in DedupBench. In this experiment, Chonkers operates on bytes, not characters. The target Chunk size is $8$~kiB based on the configurations predefined by DedupBench, but the actual average chunk size achieved by the different algorithms varies. If the last layer's absolute unit is $8$~kiB, the chunks produced by Chonkers will be significantly smaller than those from most other methods. Smaller chunks means more opportunities for deduplication, so \we employ Chonkers with an ultimate absolute unit of $12$~kiB, which leads to chunk sizes that are more meaningfully comparable.

In the results, shown in Table \vref{dedupBench}, Chonkers stands out as having by far the smallest chunk size variance, if we ignore AE-Min which has less variance, but is really bad at deduplicating. However, when it comes to the deduplication ratio, Chonkers only outperforms three of the other methods, though this comparison is not entirely fair because two of the algorithms that are better deduplicators than Chonkers also produce smaller chunks on average.

\section{Potential Applications}
Chonkers can be used for all the usual applications of content-defined chunking algorithms, such as distributed storage systems, deduplicated backups, substring searches, genomic sequence storage, and linguistic data processing. Its aggressive deduplication especially benefits monoidal reductions and versioned file systems.

Beyond these applications, Chonkers enables\footnote{Indeed, the initial motivation for inventing it was this application} finding common substrings with less memory than usual, given a lower bound for the lengths of substrings we want to find with certainty. The usual approach to finding common substrings uses suffix tries\cite{crochemore2016linear}, but these need several times more memory than the original strings. Thanks to Chonkers' weight and locality guarantees, we can choose an absolute unit small enough so that any common substrings that reach the desired minimum length will have at least one chunk\footnote{Or segment of a caterpillar} in common. Such a common chunk already represents a common substring, but is maybe only part of a longer match. The same strategy used for computing diffbits can be employed to expand the match efficiently to its maximum length.

\section{Conclusion and Future Directions}
Chonkers provides robust theoretical guarantees previously unattainable simultaneously. Future work includes exploring throughput-optimized combinations with existing chunking algorithms, leveraging aggressive deduplication for monoidal computations, and optimizing the details of  compactly storing a chonker tree in working memory as well as on block-based storage devices. These ideas are outlined in section \vref{sectionTweaks}.

\bibliographystyle{plain}
\bibliography{chonkersFoiks}

\pagebreak
\appendix

\section{Proofs of the Theoretical Guarantees}
\label{sectionProofs}
This section contains the proofs for the weight and locality guarantees.

\subsection{Strict Weight Guarantee}

\subsubsection{In the Balancing Phase}
Per the precondition, the input to the balancing phase contains no two consecutive kittens. This means that all neighbors of a kitten have weight $\ge \frac14$. But the kitten itself is by definition lighter than $\frac14$, and therefore lighter than its neighbors.
Hence, the balancing phase will attempt to merge it with its neighbors.

If the kitten is not merged with its right neighbor at priority $0$, the only possible reasons for this are that
\begin{itemize}
 \item the right neighbor does not exist.
 \item the right neighbor is not heckin' with the kitten: Prevented by the no-megachonkers rule. 
\end{itemize}
It is not possible that the merger is prevented by the determinism rule, \ie, that the right boundary of the neighbor also has merge priority $0$. This merge priority is only given to right boundaries of chunks that are lighter than all their neighbors, but as already established the neighbor is not lighter than the kitten.

If the kitten also is not merged with its left neighbor at priority $1$, then either
\begin{itemize}
\item The left neighbor does not exist.
\item The kitten is not (or, after performing the priority $0$ mergers, no longer) heckin' with the left neighbor: Prevented by the no-megachonkers rule.
\end{itemize}
It is not possible that the merger is prevented by the determinism rule, \ie, that the right boundary of the kitten also has merge priority $1$, because that boundary has merge priority $0$ if it exists.

Hence, any kitten coming out of the balancing phase is not heckin' with any of its neighbors, and therefore not a heckin' chonker, thus fulfilling the postcondition of the balancing phase.

\subsubsection{In the Caterpillar Phase}
The caterpillar phase does not cause chunks that were not heckin' in its input to be heckin' in its output. That would require increasing the absolute unit, or splitting a chunk, neither of which happens.

The caterpillar phase does not generate any new kittens. It only generates caterpillars, and for a caterpillar to be a kitten, its constituents would have to be consecutive kittens, which is excluded by the precondition because they would be heckin' chonkers.

Mergers only increase weight: a run of two or more equal chunks has total weight at least the weight of one constituent, so it cannot yield a chunk of weight $<\tfrac14$ unless there were two consecutive kittens, which the precondition excludes.

Hence if the input to the caterpillar phase contains no kittens that are heckin' chonkers, then neither does its output, and the postcondition of the balancing phase is preserved.

The caterpillar phase can only assemble caterpillars from chunks present in its input. All of these will be lighter than the absolute unit, unless 
\begin{itemize}
\item They are caterpillars formed in a previous layer. But those would already be maximal, so consecutive such caterpillars would be unequal and would not be combined into an even bigger caterpillar.
\item They were already present in the proto-chunking. In that case, we can forego turning them into a caterpillar in order to keep the promise that caterpillars have segments lighter than the absolute unit. But this is not important.
\end{itemize}

\subsubsection{In the Diffbit Phase}
By the precondition, there are no kittens that are heckin' chonkers. 

Consequently, all runs of two or more consecutive chunks where each is heckin' with its neighbors in the run, do not contain any chunks lighter than $\tfrac14$. This means that merging any two neighboring chunks in such a run must yield a heftychonk and not a fine boi.

So, if there were two consecutive fine bois $l$ and $r$ to come out of the diffbit phase, neither of them would have undergone a merger.
But that is not possible: let $p$ be the merge priority ascribed to the boundary between $l$ and $r$, with neither $l$ nor $r$ being the product of a merger in the current diffbit phase. The right boundary of the right fine boi $r$ cannot have merge priority $p$: The initially ascribed merge priorities, being diffbits, are distinct, and the boundaries of $r$ still carry the original merge priorities because by assumption $r$ was not merged with anything so far. The merger is thus not prevented by the determinism rule. Moreover, $l$ and $r$ would be heckin' with each other because both are lighter than $\tfrac12$, and so together they are lighter than $1$ absolute unit. Hence, the merger is not prevented by the no-megachonkers rule. Nothing prevents the two fine bois from being merged at priority $p$.

In conclusion, any fine boi that comes out of the diffbit phase will be isolated and not next to another fine boi or a kitten.

Consecutive kittens are also impossible, as there were already none in the input.

Thus, the postcondition is met.

\subsubsection{Assembling the Proof}
The proof now works by induction over the layers.

By choosing the absolute unit of the first layer light enough, the precondition for its balancing phase can always be met.

By increasing the absolute unit between layers by no more than a factor of $2$, some or all fine bois will be reclassified as kittens, but no heftychonk or megachonker would be reclassified as a kitten. The postcondition of the diffbit phase of the previous layer thus implies the precondition of the balancing phase of the current layer, enabling the proof by induction.

Once the absolute unit has increased to the desired target maximum weight, it is now easy to see that the following guarantees hold:
\begin{itemize}
\item megachonkers, that is chunks that reach or exceed the target weight, can only occur due to overweight proto-chunks or caterpillars of period $< 1$
\item Of any two consecutive chunks, only one can be lighter than half the target weight (a fine boi or a kitten). 
\item Moreover, if a chunk is lighter than a quarter of the target weight (a kitten), then the sum of its weight and any of its neighbors must reach or exceed the target weight. This is because it must have been a kitten already in the input of the last layer, and can therefore not be heckin' with any of its neighbors due to the balancing phase postcondition.
\item With these conditions, the lowest possible\footnote{Whether this weight sequence can actually be produced by Chonkers is irrelevant; \We only want to construct the worst possible sequence compatible with the stated weight guarantees. A more detailed analysis should lead to a tighter bound here.} average weight when excluding the last or first chunk (whichever is lighter) is reached by alternating chunks of weight $\tfrac14$ and weight $\tfrac12$. The average weight is thus at least $\tfrac38$ of the target maximum weight, modulo effects of the ends of the sequence.
\end{itemize}

\subsection{Strict Locality Guarantee}
There are several ways in which the merge decisions can be affected by data that is farther away:
\begin{itemize}
\item Balancing phase merge priorities
\item Diffbit computation
\item Cascade effects due to the no-megachonkers and determinism rules
\item Compounding influence from previous layers
\end{itemize}

\subsubsection{Balancing Phase Merge Priorities}
The merge priorities ascribed to chunk boundaries here can be influenced by up to two chunks in both directions:
Without loss of generality, let us assume that chunk $l$ is to the left of chunk $k$ and heavier than $k$. 
Then the boundary between them might be assigned the merge priority $1$, but only if $k$ is also lighter than its right neighbor $r$.
Thus, the weight of chunk $r$ has an influence on the decision to merge chunk $l$ that is two chunks to its left.
The influence on the two chunks to the right works analogously.
Because $k$ has to be heckin' with both its neighbors for the influence to spread across it, the total weight of the chunks involved cannot exceed $2$ absolute units in either direction.

\subsubsection{Causal Influence of Diffbit Computation}
Let $b_o$ be the maximum weight of augmented bit content for which $o$ orders of diffbits are needed to arrive at a highest order diffbit between $0$ and $5$. It is easy to see that the sequence $b$ is given by
\begin{eqnarray}
b_1 & = 3 \\
b_{o+1} & = 2^{b_o - 1},\\
b &= \left(3, 4, 8, 128, 2^{127}\mathellipsis \right).
\end{eqnarray}
This is very similar to iterated exponentiation defined as
\begin{eqnarray}
e_1 & = 2 \\
e_{o+1} & = 2^{e_o} \\
e &= \left(2, 4, 16, 65536, 2^{65536}\mathellipsis \right),
\end{eqnarray}
yet due to the subtraction of $1$ in the exponent, $b_o \leq e_o$ except for $o=1$. However, it is easy to see that $b_{o+1}>e_o$.
This implies that the difference of the inverses\footnote{For values that do not occur in the original sequence, the value of the inverse sequence can be taken to be the $o$ value belonging to the next value that does occur.} of $e$ and $o$ is bounded by a constant, and so the number of diffbit orders needed to assign a merge priority between $0$ and $5$ to chunks with raw bit content weight up to $\au$ grows as $\Theta(\log^* \au)$.

A change somewhere in the data, even if it would not affect the chunk boundaries coming out of previous layers, can affect the diffbits of the chunks to the right, and of the chunk that contains the change. The highest order diffbit is affected by $O(\log^* \au)$ chunks, which is in practice never more than seven consecutive chunks. One of these chunks is to the left of the boundary carrying the highest order diffbit as its priority, and the rest is to the right. Since diffbit computation does not propagate information over chunk boundaries that are not heck'd, all these chunks need to be heckin' with their neighbors in order for this worst case to be realized. 

The maximum total weight of such a sequence of seven heckin' chonkers is realized by an alternating sequence of maximally light and maximally heavy chunks, starting with a heavy chunk, so that any two neighbors in the sequence weigh just below the absolute unit together. Even if we allow the light chunks to have a weight of $1$ bit, the total weight of the alternating sequence cannot exceed $4$ absolute units. Due to the weight guarantees, it is even somewhat less than that, but \we will not make use of this in the proof of the locality guarantees.

\subsubsection{Causal Influence Via Cascade Effects}
The weight and merge priorities of neighboring chunks can affect the decision to merge two chunks. This, in turn, can affect their neighbors.
The priority-based merge rules are designed to make merging decisions as local as possible, but certain cascading effects cannot be avoided.
A decision whether to merge a chunk boundary at priority $p$ is influenced by whether one of the involved chunks has already been merged with its other neighbor at an earlier priority: Such a merger might trigger the no-megachonkers rule because the combined chunk is now too big. Additionally, if the right chunk of a prospective merger has already undergone a merger at an earlier priority, the right boundary of the result might also be annotated with priority $p$, preventing the merger of the two chunks to its left according to the determinism rule.

To illustrate, consider the following sequence of chunks, represented by boxes with diffbit-based merge priorities written between them:
$$ \Box \: 0 \: \Box  \: 1  \: \Box \:  2 \:  \Box \:  3 \:  \Box  \: 4 \:  \Box  \: 5  \: \Box $$
If the chunk weights are such that the result of merging any two neighbors is no longer a heckin' chonker, then this will get merged as follows:
$$ \boxed{\Box 0 \Box} \:  1 \:  \boxed{\Box 2 \Box}  \: 3 \:  \boxed{\Box 4 \Box}  \: 5 \:  \Box$$
This is because the no-megachonkers rule prevents every second possible merger.
Now observe the effect of changing the leftmost merge priority:
$$ \Box  \: 3 \:  \Box  \: 1 \:  \Box \:  2  \: \Box \:  3 \:  \Box \:  4 \:  \Box \:  5  \: \Box $$
will get merged as
$$ \Box \:  3  \: \boxed{\Box 1 \Box}  \: 2  \: \boxed{\Box 3 \Box} \:  4 \:  \boxed{\Box 5 \Box} $$
A cascade can involve even more chunks in the input of the priority-based merging process, which then get assembled into the cascading chunks during the merge process. The following diagrams illustrate this:
$$ \Box  \: 0  \: \Box  \: 1   \:        \Box  \: 0  \: \Box \:  2 \:  \Box \: 3   \:       \Box \:  0 \:  \Box  \:  4  \: \Box  \: 5  \: \Box \:  0  \: \Box $$
Would get merged to yield the intermediate state
$$ \Box \:  0  \: \Box \:  1 \:  \boxed{\Box 0 \Box}  \: 2 \:  \Box \:  3  \: \boxed{\Box 0 \Box} \:  4  \: \Box \:  5  \: \boxed{\Box 0 \Box}, $$
which then could possibly exhibit the cascade described above. 
While such a cascade can extend for more than seven chunks, these would have to be smaller so that the no-megachonkers rule does not interrupt the cascade.
Therefore, the total weight of chunks taking part in a cascade cannot exceed $4$ absolute units, as discussed above in the case of causal influences extending for seven chunks in the case of diffbit computation.

A cascade can also be formed by a descending sequence of merge priorities, thus extending the causal influence to the left. One might think that the determinism rule could be used to construct a cascade that is even longer by one chunk. But a maximally long descending no-megachonkers-based cascade will have $0$ as its rightmost merge priority. When merging the chunks at priority $0$, we have the guarantee that consecutive priorities are different because they are diffbits\footnote{This is speaking for the diffbit phase. In the balancing phase, it is also easy to see why they are different: Priority $0$ is only used at the right boundary, and priority $1$ only at the left boundary, of locally lightest chunks, which cannot occur next to each other} and no chunk boundaries have yet been eliminated by earlier mergers. Thus the rightmost merger can actually not be affected by the determinism rule if the cascade involves priorities descending to $0$. If the no-megachonkers cascade is shorter, the weight of the affected blocks can increase by less than one absolute unit by virtue of the determinism rule, but this does not affect the worst case.

Another way to see this is that equal consecutive merge priorities $p$ at the right boundary of a cascade can only happen if the right chunk of the two with priority $p$ at its right boundary has arisen from merging two chunks at an earlier priority. Before that merger, we had just an ordinary cascade that did not involve the determinism rule. Hence, to account for the effect of cascades, the determinism rule is unnecessary.

The cascades in the balancing phase work the same way, but their causal influence is limited to $2$ absolute units in both directions because they involve only $2$ merge priorities and $3$ relevant heckin' chonkers.

In conclusion, the causal influence due to cascades within a single layer reaches up to $6$ absolute units in both directions.

\subsubsection{Compounding Influence from Previous Layers}
Together with the causal influence due to the way merge priorities are defined, merge priorities are affected by the input to the layer for at most $(6+2+1)=9$ absolute units to the right and $(6+2+4)=12$ absolute units to the left. 

But the chunk weights, chunk contents, and thus the diffbits, used for the merge decisions in a layer, are affected by the decisions in the previous layers. Assuming the absolute unit doubles between layers, the worst case causal influence from the previous layer will only extend for half the number of bits as that of the current layer, and the influence from the layer before that only for a quarter of the number of bits and so on. These influence ranges add, because any change to the chunk structure going into a layer can change the layer's output for as far as that layer propagates influences. The compounded influence of all layers is thus bounded to be larger than the influence of the highest layer by a factor given by the geometric series $\sum_{i=0}^\infty 2^{-i} = 2$.

In conclusion, a single change in the underlying bit sequence can affect the chunking for up to $18$ absolute units to the right and up to $24$ absolute units to the left. But that is only an easily provable upper bound; the actual worst case might be better due to interactions between parts of the algorithm not considered in the derivation. Furthermore, the average case will be much better still as the worst case requires very specific arrangements of chunk weights and diffbits. For an experimental evaluation of the actual behavior of Chonkers, see section \vref{xp}.

If the ultimate absolute unit of the final layer is not a multiple of the initial absolute unit by a power of $2$, then the locality bound will be weaker because the compounding influence of previous layers can exceed the bound derived from the geometric series. To strengthen the bound, the larger deviations from doubling the absolute unit between layers should occur early, so that the excess influence they contribute is less when measured relative to the ultimate absolute unit.

\pagebreak

\FloatBarrier

\section{Experimental Results: Data}

\begin{table}[h]
\center
{
\small
\begin{tabular}{r|r|r|r|r|}
\ccnf{layer} & \cc{average} & \cc{$\sigma$} & \cc{max segment} & \cc{min pair} \\\hline
$1$ & $0.89 \pm 0.09 \in [0.75; 2.95]$ & $0.359 \pm 0.501$ & $0.98 \pm 0.00 \in [0.98; 0.98]$ & $1.78 \pm 0.18 \in [1.50; 5.93]$ \\
$2$ & $0.79 \pm 0.08 \in [0.65; 2.54]$ & $0.266 \pm 0.313$ & $0.99 \pm 0.00 \in [0.74; 0.99]$ & $1.57 \pm 0.15 \in [1.32; 5.12]$ \\
$3$ & $0.75 \pm 0.06 \in [0.62; 2.01]$ & $0.225 \pm 0.187$ & $1.00 \pm 0.00 \in [0.75; 1.00]$ & $1.50 \pm 0.13 \in [1.21; 4.07]$ \\
$4$ & $0.73 \pm 0.05 \in [0.54; 2.20]$ & $0.205 \pm 0.158$ & $1.00 \pm 0.01 \in [0.56; 1.00]$ & $1.45 \pm 0.10 \in [1.08; 4.41]$ \\
$5$ & $0.71 \pm 0.04 \in [0.49; 1.67]$ & $0.187 \pm 0.096$ & $1.00 \pm 0.02 \in [0.56; 1.00]$ & $1.42 \pm 0.07 \in [1.03; 3.34]$ \\
$6$ & $0.70 \pm 0.03 \in [0.49; 1.36]$ & $0.180 \pm 0.068$ & $0.99 \pm 0.03 \in [0.53; 1.00]$ & $1.40 \pm 0.06 \in [1.02; 2.74]$ \\
$7$ & $0.70 \pm 0.04 \in [0.46; 1.17]$ & $0.176 \pm 0.052$ & $0.98 \pm 0.05 \in [0.52; 1.00]$ & $1.40 \pm 0.07 \in [1.01; 2.34]$ \\
$8$ & $0.69 \pm 0.05 \in [0.44; 1.17]$ & $0.173 \pm 0.043$ & $0.96 \pm 0.07 \in [0.50; 1.00]$ & $1.39 \pm 0.09 \in [1.00; 2.35]$ \\
$9$ & $0.69 \pm 0.05 \in [0.44; 1.04]$ & $0.170 \pm 0.042$ & $0.95 \pm 0.08 \in [0.51; 1.00]$ & $1.38 \pm 0.11 \in [1.00; 2.08]$ \\
$10$ & $0.69 \pm 0.06 \in [0.44; 2.30]$ & $0.164 \pm 0.050$ & $0.93 \pm 0.09 \in [0.50; 1.00]$ & $1.38 \pm 0.13 \in [1.00; 4.92]$ \\
$11$ & $0.68 \pm 0.07 \in [0.43; 1.72]$ & $0.161 \pm 0.053$ & $0.90 \pm 0.10 \in [0.50; 1.00]$ & $1.37 \pm 0.15 \in [1.00; 3.82]$ \\
$12$ & $0.68 \pm 0.08 \in [0.44; 1.38]$ & $0.156 \pm 0.059$ & $0.88 \pm 0.11 \in [0.50; 1.00]$ & $1.36 \pm 0.16 \in [1.00; 3.19]$ \\
$13$ & $0.67 \pm 0.09 \in [0.45; 1.00]$ & $0.150 \pm 0.064$ & $0.86 \pm 0.12 \in [0.47; 1.00]$ & $1.35 \pm 0.18 \in [1.00; 1.99]$ \\
$14$ & $0.67 \pm 0.09 \in [0.44; 0.99]$ & $0.146 \pm 0.069$ & $0.84 \pm 0.12 \in [0.50; 1.00]$ & $1.34 \pm 0.19 \in [1.00; 1.99]$ \\
$15$ & $0.66 \pm 0.10 \in [0.45; 0.99]$ & $0.144 \pm 0.072$ & $0.82 \pm 0.13 \in [0.50; 1.00]$ & $1.32 \pm 0.19 \in [1.00; 1.98]$ \\
$16$ & $0.64 \pm 0.10 \in [0.44; 0.99]$ & $0.141 \pm 0.072$ & $0.80 \pm 0.13 \in [0.51; 1.00]$ & $1.29 \pm 0.19 \in [1.00; 1.99]$ \\
$17$ & $0.65 \pm 0.09 \in [0.46; 0.92]$ & $0.143 \pm 0.067$ & $0.82 \pm 0.13 \in [0.51; 1.00]$ & $1.30 \pm 0.18 \in [1.00; 1.84]$ \\
$18$ & $0.66 \pm 0.08 \in [0.47; 0.92]$ & $0.151 \pm 0.065$ & $0.85 \pm 0.13 \in [0.54; 1.00]$ & $1.33 \pm 0.17 \in [1.00; 1.84]$ \\
$19$ & $0.68 \pm 0.08 \in [0.50; 0.97]$ & $0.153 \pm 0.057$ & $0.87 \pm 0.11 \in [0.60; 1.00]$ & $1.35 \pm 0.16 \in [1.03; 1.95]$ \\
$20$ & $0.67 \pm 0.09 \in [0.49; 0.91]$ & $0.159 \pm 0.065$ & $0.87 \pm 0.12 \in [0.52; 1.00]$ & $1.35 \pm 0.17 \in [1.00; 1.84]$ \\
$21$ & $0.69 \pm 0.10 \in [0.50; 0.96]$ & $0.142 \pm 0.056$ & $0.87 \pm 0.12 \in [0.59; 1.00]$ & $1.37 \pm 0.21 \in [1.00; 1.89]$ \\
$22$ & $0.67 \pm 0.09 \in [0.51; 0.84]$ & $0.144 \pm 0.081$ & $0.83 \pm 0.12 \in [0.52; 0.99]$ & $1.34 \pm 0.17 \in [1.01; 1.63]$ \\
$23$ & $0.68 \pm 0.11 \in [0.52; 0.93]$ & $0.161 \pm 0.092$ & $0.84 \pm 0.12 \in [0.68; 0.99]$ & $1.38 \pm 0.20 \in [1.04; 1.86]$ \\
$24$ & $0.71 \pm 0.00 \in [0.71; 0.71]$ & $0.242 \pm 0.000$ & $0.96 \pm 0.00 \in [0.96; 0.96]$ & $1.43 \pm 0.00 \in [1.43; 1.43]$ \\
\hline
\end{tabular}
}

\caption{Weight statistics for the kernel source code corpus}
\label{kernelsize}
\end{table}

\begin{table}
\center
{
\small
\begin{tabular}{r|r|r|r|r|}
\ccnf{layer} & \cc{average} & \cc{$\sigma$} & \cc{max segment} & \cc{min pair} \\\hline
$1$ & $0.87 \pm 0.01 \in [0.85; 1.27]$ & $0.250 \pm 0.097$ & $0.98 \pm 0.00 \in [0.98; 0.98]$ & $1.74 \pm 0.02 \in [1.69; 2.54]$ \\
$2$ & $0.77 \pm 0.01 \in [0.75; 1.09]$ & $0.209 \pm 0.055$ & $0.99 \pm 0.00 \in [0.99; 0.99]$ & $1.54 \pm 0.02 \in [1.51; 2.19]$ \\
$3$ & $0.73 \pm 0.01 \in [0.71; 0.98]$ & $0.193 \pm 0.029$ & $1.00 \pm 0.00 \in [1.00; 1.00]$ & $1.47 \pm 0.01 \in [1.41; 1.97]$ \\
$4$ & $0.72 \pm 0.00 \in [0.69; 0.84]$ & $0.187 \pm 0.013$ & $1.00 \pm 0.00 \in [1.00; 1.00]$ & $1.44 \pm 0.01 \in [1.38; 1.67]$ \\
$5$ & $0.71 \pm 0.00 \in [0.67; 0.73]$ & $0.182 \pm 0.005$ & $1.00 \pm 0.00 \in [0.94; 1.00]$ & $1.41 \pm 0.01 \in [1.34; 1.46]$ \\
$6$ & $0.70 \pm 0.00 \in [0.63; 0.75]$ & $0.180 \pm 0.003$ & $1.00 \pm 0.00 \in [0.94; 1.00]$ & $1.40 \pm 0.01 \in [1.27; 1.49]$ \\
$7$ & $0.70 \pm 0.01 \in [0.62; 0.77]$ & $0.179 \pm 0.004$ & $1.00 \pm 0.00 \in [0.85; 1.00]$ & $1.40 \pm 0.01 \in [1.21; 1.53]$ \\
$8$ & $0.70 \pm 0.01 \in [0.57; 0.84]$ & $0.179 \pm 0.007$ & $1.00 \pm 0.01 \in [0.71; 1.00]$ & $1.39 \pm 0.02 \in [1.21; 1.68]$ \\
$9$ & $0.70 \pm 0.01 \in [0.52; 0.84]$ & $0.178 \pm 0.009$ & $1.00 \pm 0.02 \in [0.59; 1.00]$ & $1.39 \pm 0.03 \in [1.03; 1.69]$ \\
$10$ & $0.70 \pm 0.02 \in [0.49; 0.91]$ & $0.178 \pm 0.012$ & $0.99 \pm 0.02 \in [0.65; 1.00]$ & $1.39 \pm 0.04 \in [1.08; 1.81]$ \\
$11$ & $0.70 \pm 0.03 \in [0.51; 0.96]$ & $0.176 \pm 0.017$ & $0.99 \pm 0.03 \in [0.55; 1.00]$ & $1.39 \pm 0.05 \in [1.02; 1.91]$ \\
$12$ & $0.69 \pm 0.03 \in [0.49; 0.94]$ & $0.175 \pm 0.024$ & $0.98 \pm 0.05 \in [0.56; 1.00]$ & $1.39 \pm 0.07 \in [1.00; 1.88]$ \\
$13$ & $0.69 \pm 0.04 \in [0.50; 0.95]$ & $0.172 \pm 0.031$ & $0.96 \pm 0.07 \in [0.53; 1.00]$ & $1.39 \pm 0.09 \in [1.00; 1.90]$ \\
$14$ & $0.69 \pm 0.05 \in [0.46; 0.96]$ & $0.169 \pm 0.037$ & $0.95 \pm 0.08 \in [0.51; 1.00]$ & $1.38 \pm 0.10 \in [1.00; 1.93]$ \\
$15$ & $0.69 \pm 0.06 \in [0.48; 0.97]$ & $0.166 \pm 0.042$ & $0.93 \pm 0.08 \in [0.51; 1.00]$ & $1.38 \pm 0.12 \in [1.00; 1.93]$ \\
$16$ & $0.69 \pm 0.07 \in [0.47; 0.96]$ & $0.163 \pm 0.048$ & $0.91 \pm 0.09 \in [0.51; 1.00]$ & $1.38 \pm 0.14 \in [1.00; 1.93]$ \\
$17$ & $0.68 \pm 0.08 \in [0.47; 0.98]$ & $0.155 \pm 0.059$ & $0.88 \pm 0.11 \in [0.52; 1.00]$ & $1.37 \pm 0.16 \in [1.00; 1.96]$ \\
$18$ & $0.67 \pm 0.10 \in [0.44; 0.97]$ & $0.147 \pm 0.071$ & $0.83 \pm 0.12 \in [0.51; 1.00]$ & $1.34 \pm 0.19 \in [1.00; 1.94]$ \\
$19$ & $0.65 \pm 0.10 \in [0.45; 0.96]$ & $0.141 \pm 0.077$ & $0.80 \pm 0.12 \in [0.52; 1.00]$ & $1.30 \pm 0.19 \in [1.00; 1.92]$ \\
$20$ & $0.62 \pm 0.10 \in [0.49; 0.97]$ & $0.126 \pm 0.083$ & $0.76 \pm 0.12 \in [0.51; 0.99]$ & $1.25 \pm 0.20 \in [1.00; 1.94]$ \\
$21$ & $0.56 \pm 0.03 \in [0.51; 0.62]$ & $0.161 \pm 0.057$ & $0.75 \pm 0.10 \in [0.52; 0.91]$ & $1.14 \pm 0.08 \in [1.01; 1.27]$ \\
$22$ & $0.60 \pm 0.02 \in [0.58; 0.62]$ & $0.109 \pm 0.068$ & $0.71 \pm 0.08 \in [0.62; 0.79]$ & $1.20 \pm 0.03 \in [1.17; 1.23]$ \\
\hline
\end{tabular}

}

\caption{Weight statistics for the German fiction corpus}
\label{fictionsize}
\end{table}

\skipAll{
\begin{table}
\center
{
\small 
\begin{tabular}{r|r|r|r|r|}
\ccnf{layer} & \cc{average} & \cc{$\sigma$} & \cc{max segment} & \cc{min pair} \\\hline
$1$ & $0.89 \pm 0.04 \in [0.80; 1.55]$ & $0.320 \pm 0.202$ & $0.98 \pm 0.00 \in [0.98; 0.98]$ & $1.78 \pm 0.08 \in [1.60; 3.10]$ \\
$2$ & $0.78 \pm 0.03 \in [0.67; 1.32]$ & $0.227 \pm 0.116$ & $0.99 \pm 0.00 \in [0.99; 0.99]$ & $1.56 \pm 0.06 \in [1.35; 2.63]$ \\
$3$ & $0.74 \pm 0.02 \in [0.61; 1.11]$ & $0.198 \pm 0.062$ & $1.00 \pm 0.00 \in [0.87; 1.00]$ & $1.48 \pm 0.04 \in [1.25; 2.23]$ \\
$4$ & $0.72 \pm 0.02 \in [0.55; 0.95]$ & $0.186 \pm 0.032$ & $1.00 \pm 0.01 \in [0.75; 1.00]$ & $1.44 \pm 0.04 \in [1.17; 1.91]$ \\
$5$ & $0.71 \pm 0.02 \in [0.48; 0.93]$ & $0.180 \pm 0.020$ & $0.99 \pm 0.03 \in [0.62; 1.00]$ & $1.41 \pm 0.05 \in [1.08; 1.89]$ \\
$6$ & $0.70 \pm 0.03 \in [0.46; 1.56]$ & $0.178 \pm 0.053$ & $0.99 \pm 0.04 \in [0.52; 1.00]$ & $1.40 \pm 0.07 \in [1.02; 3.17]$ \\
$7$ & $0.70 \pm 0.04 \in [0.45; 1.65]$ & $0.176 \pm 0.041$ & $0.98 \pm 0.06 \in [0.52; 1.00]$ & $1.39 \pm 0.08 \in [1.01; 3.40]$ \\
$8$ & $0.69 \pm 0.05 \in [0.45; 1.46]$ & $0.173 \pm 0.038$ & $0.96 \pm 0.07 \in [0.52; 1.00]$ & $1.39 \pm 0.09 \in [1.01; 3.06]$ \\
$9$ & $0.69 \pm 0.05 \in [0.44; 1.17]$ & $0.169 \pm 0.038$ & $0.95 \pm 0.08 \in [0.50; 1.00]$ & $1.39 \pm 0.11 \in [1.00; 2.59]$ \\
$10$ & $0.69 \pm 0.06 \in [0.45; 0.99]$ & $0.165 \pm 0.042$ & $0.93 \pm 0.09 \in [0.51; 1.00]$ & $1.38 \pm 0.12 \in [1.00; 2.00]$ \\
$11$ & $0.69 \pm 0.07 \in [0.44; 0.98]$ & $0.164 \pm 0.048$ & $0.92 \pm 0.10 \in [0.50; 1.00]$ & $1.38 \pm 0.14 \in [1.00; 1.97]$ \\
$12$ & $0.68 \pm 0.07 \in [0.46; 0.97]$ & $0.161 \pm 0.054$ & $0.91 \pm 0.11 \in [0.51; 1.00]$ & $1.37 \pm 0.14 \in [1.00; 1.94]$ \\
$13$ & $0.68 \pm 0.07 \in [0.46; 0.97]$ & $0.157 \pm 0.054$ & $0.88 \pm 0.11 \in [0.50; 1.00]$ & $1.35 \pm 0.15 \in [1.00; 1.95]$ \\
$14$ & $0.66 \pm 0.09 \in [0.45; 0.98]$ & $0.157 \pm 0.056$ & $0.86 \pm 0.11 \in [0.51; 1.00]$ & $1.34 \pm 0.18 \in [1.00; 1.95]$ \\
$15$ & $0.69 \pm 0.11 \in [0.49; 0.99]$ & $0.142 \pm 0.074$ & $0.85 \pm 0.11 \in [0.50; 1.00]$ & $1.38 \pm 0.21 \in [1.00; 1.97]$ \\
$16$ & $0.68 \pm 0.10 \in [0.47; 0.96]$ & $0.138 \pm 0.082$ & $0.83 \pm 0.10 \in [0.52; 1.00]$ & $1.35 \pm 0.19 \in [1.01; 1.93]$ \\
$17$ & $0.64 \pm 0.09 \in [0.50; 0.89]$ & $0.158 \pm 0.076$ & $0.80 \pm 0.14 \in [0.56; 1.00]$ & $1.27 \pm 0.19 \in [1.01; 1.77]$ \\
$18$ & $0.73 \pm 0.10 \in [0.56; 0.83]$ & $0.105 \pm 0.057$ & $0.84 \pm 0.10 \in [0.65; 0.98]$ & $1.47 \pm 0.19 \in [1.12; 1.66]$ \\
$19$ & $0.69 \pm 0.06 \in [0.58; 0.76]$ & $0.110 \pm 0.061$ & $0.83 \pm 0.12 \in [0.68; 0.98]$ & $1.40 \pm 0.11 \in [1.21; 1.57]$ \\
$20$ & $0.61 \pm 0.10 \in [0.53; 0.76]$ & $0.100 \pm 0.028$ & $0.71 \pm 0.12 \in [0.60; 0.87]$ & $1.23 \pm 0.23 \in [1.06; 1.55]$ \\
$21$ & $0.76 \pm 0.00 \in [0.76; 0.76]$ & $0.088 \pm 0.000$ & $0.85 \pm 0.00 \in [0.85; 0.85]$ & $1.52 \pm 0.00 \in [1.52; 1.52]$ \\
\hline
\end{tabular}
}

\caption{Weight statistics for the Java source code corpus}
\label{javasize}
\end{table}
}

\begin{table}
\center
{
\small
\begin{tabular}{r|r|r|r|r|}
\ccnf{layer} & \cc{average} & \cc{$\sigma$} & \cc{max segment} & \cc{min pair} \\\hline
$1$ & $0.87 \pm 0.00 \in [0.86; 0.87]$ & $0.211 \pm 0.001$ & $0.98 \pm 0.00 \in [0.98; 0.98]$ & $1.73 \pm 0.00 \in [1.72; 1.75]$ \\
$2$ & $0.77 \pm 0.00 \in [0.76; 0.78]$ & $0.186 \pm 0.001$ & $0.99 \pm 0.00 \in [0.99; 0.99]$ & $1.54 \pm 0.01 \in [1.52; 1.56]$ \\
$3$ & $0.74 \pm 0.00 \in [0.72; 0.75]$ & $0.184 \pm 0.002$ & $1.00 \pm 0.00 \in [1.00; 1.00]$ & $1.47 \pm 0.01 \in [1.45; 1.51]$ \\
$4$ & $0.72 \pm 0.01 \in [0.70; 0.74]$ & $0.182 \pm 0.003$ & $1.00 \pm 0.00 \in [1.00; 1.00]$ & $1.43 \pm 0.01 \in [1.40; 1.48]$ \\
$5$ & $0.71 \pm 0.01 \in [0.68; 0.74]$ & $0.180 \pm 0.004$ & $1.00 \pm 0.00 \in [1.00; 1.00]$ & $1.41 \pm 0.01 \in [1.36; 1.48]$ \\
$6$ & $0.70 \pm 0.01 \in [0.66; 0.74]$ & $0.179 \pm 0.006$ & $1.00 \pm 0.00 \in [0.98; 1.00]$ & $1.40 \pm 0.02 \in [1.32; 1.49]$ \\
$7$ & $0.70 \pm 0.01 \in [0.65; 0.76]$ & $0.178 \pm 0.008$ & $1.00 \pm 0.00 \in [0.95; 1.00]$ & $1.40 \pm 0.03 \in [1.29; 1.52]$ \\
$8$ & $0.70 \pm 0.02 \in [0.63; 0.78]$ & $0.178 \pm 0.012$ & $0.99 \pm 0.01 \in [0.91; 1.00]$ & $1.40 \pm 0.04 \in [1.26; 1.56]$ \\
$9$ & $0.70 \pm 0.03 \in [0.61; 0.81]$ & $0.176 \pm 0.017$ & $0.98 \pm 0.02 \in [0.84; 1.00]$ & $1.39 \pm 0.06 \in [1.22; 1.64]$ \\
$10$ & $0.70 \pm 0.04 \in [0.57; 0.89]$ & $0.173 \pm 0.025$ & $0.96 \pm 0.04 \in [0.75; 1.00]$ & $1.39 \pm 0.08 \in [1.14; 1.79]$ \\
$11$ & $0.69 \pm 0.06 \in [0.54; 0.81]$ & $0.168 \pm 0.038$ & $0.93 \pm 0.06 \in [0.62; 1.00]$ & $1.39 \pm 0.12 \in [1.08; 1.74]$ \\
$12$ & $0.70 \pm 0.10 \in [0.49; 0.81]$ & $0.143 \pm 0.057$ & $0.87 \pm 0.09 \in [0.62; 1.00]$ & $1.40 \pm 0.20 \in [1.02; 1.72]$ \\
$13$ & $0.61 \pm 0.00 \in [0.61; 0.61]$ & $0.146 \pm 0.086$ & $0.76 \pm 0.09 \in [0.61; 1.00]$ & $1.22 \pm 0.00 \in [1.22; 1.22]$ \\
\hline
\end{tabular} 
}

\caption{Weight statistics for the random corpus}
\label{randomsize}
\end{table}


\begin{table}
\center
\begin{tabular}{r|r|r|}
\ccnf{layer} & \cc{left} & \cc{right} \\
\hline
$1$ & $0.079 \pm 0.239 \leq 2.9538$ & $0.458 \pm 0.487 \leq 2.4615$ \\
$2$ & $0.152 \pm 0.329 \leq 3.4729$ & $0.582 \pm 0.515 \leq 2.7287$ \\
$3$ & $0.189 \pm 0.376 \leq 3.9844$ & $0.603 \pm 0.544 \leq 3.1128$ \\
$4$ & $0.205 \pm 0.374 \leq 4.0546$ & $0.567 \pm 0.545 \leq 3.3060$ \\
$5$ & $0.206 \pm 0.373 \leq 4.0585$ & $0.516 \pm 0.540 \leq 3.2468$ \\
$6$ & $0.197 \pm 0.367 \leq 4.1542$ & $0.466 \pm 0.533 \leq 3.2016$ \\
$7$ & $0.183 \pm 0.355 \leq 3.8975$ & $0.412 \pm 0.517 \leq 3.1242$ \\
$8$ & $0.162 \pm 0.333 \leq 3.5582$ & $0.359 \pm 0.495 \leq 3.1871$ \\
$9$ & $0.146 \pm 0.320 \leq 4.0994$ & $0.310 \pm 0.470 \leq 3.2029$ \\
$10$ & $0.125 \pm 0.293 \leq 4.2225$ & $0.263 \pm 0.439 \leq 3.0448$ \\
$11$ & $0.106 \pm 0.269 \leq 3.8393$ & $0.221 \pm 0.405 \leq 3.0229$ \\
$12$ & $0.088 \pm 0.241 \leq 3.8803$ & $0.181 \pm 0.364 \leq 3.0693$ \\
$13$ & $0.073 \pm 0.216 \leq 3.0035$ & $0.147 \pm 0.326 \leq 3.0907$ \\
$14$ & $0.062 \pm 0.194 \leq 2.9503$ & $0.122 \pm 0.293 \leq 2.9643$ \\
$15$ & $0.052 \pm 0.173 \leq 2.7213$ & $0.100 \pm 0.258 \leq 2.6943$ \\
$16$ & $0.045 \pm 0.161 \leq 2.7082$ & $0.089 \pm 0.241 \leq 2.7143$ \\
$17$ & $0.051 \pm 0.178 \leq 2.2142$ & $0.099 \pm 0.265 \leq 2.5567$ \\
$18$ & $0.057 \pm 0.200 \leq 2.0492$ & $0.119 \pm 0.304 \leq 2.3734$ \\
$19$ & $0.071 \pm 0.204 \leq 1.6444$ & $0.130 \pm 0.312 \leq 2.0830$ \\
$20$ & $0.062 \pm 0.201 \leq 1.7236$ & $0.112 \pm 0.284 \leq 2.1687$ \\
$21$ & $0.079 \pm 0.236 \leq 1.6587$ & $0.107 \pm 0.276 \leq 2.1097$ \\
$22$ & $0.056 \pm 0.201 \leq 1.7532$ & $0.072 \pm 0.208 \leq 1.4747$ \\
$23$ & $0.070 \pm 0.232 \leq 1.4523$ & $0.063 \pm 0.182 \leq 1.1561$ \\
$24$ & $0.128 \pm 0.193 \leq 0.5275$ & $0.000 \pm 0.000 \leq 0.0000$ \\
\hline
\end{tabular}

\caption{Locality statistics for the kernel source code corpus}
\label{kernellocality}
\end{table}

\begin{table}
\center
\begin{tabular}{r|r|r|}
\ccnf{layer} & \cc{left} & \cc{right} \\
\hline
$1$ & $0.088 \pm 0.248 \leq 2.4615$ & $0.444 \pm 0.482 \leq 1.9692$ \\
$2$ & $0.162 \pm 0.317 \leq 2.7287$ & $0.609 \pm 0.527 \leq 2.4806$ \\
$3$ & $0.209 \pm 0.386 \leq 3.1128$ & $0.639 \pm 0.541 \leq 2.8638$ \\
$4$ & $0.227 \pm 0.398 \leq 3.1813$ & $0.605 \pm 0.550 \leq 3.0565$ \\
$5$ & $0.223 \pm 0.389 \leq 4.0273$ & $0.550 \pm 0.544 \leq 2.8098$ \\
$6$ & $0.211 \pm 0.381 \leq 3.1859$ & $0.498 \pm 0.541 \leq 2.8424$ \\
$7$ & $0.201 \pm 0.371 \leq 3.0930$ & $0.449 \pm 0.535 \leq 2.8821$ \\
$8$ & $0.189 \pm 0.364 \leq 3.3629$ & $0.408 \pm 0.521 \leq 2.9567$ \\
$9$ & $0.176 \pm 0.357 \leq 3.1424$ & $0.367 \pm 0.508 \leq 2.9529$ \\
$10$ & $0.163 \pm 0.347 \leq 3.2372$ & $0.336 \pm 0.498 \leq 2.6083$ \\
$11$ & $0.149 \pm 0.335 \leq 3.6308$ & $0.304 \pm 0.483 \leq 2.8960$ \\
$12$ & $0.134 \pm 0.312 \leq 3.1091$ & $0.274 \pm 0.463 \leq 2.9819$ \\
$13$ & $0.121 \pm 0.298 \leq 3.0688$ & $0.246 \pm 0.443 \leq 2.8834$ \\
$14$ & $0.115 \pm 0.296 \leq 3.1425$ & $0.220 \pm 0.419 \leq 2.8768$ \\
$15$ & $0.105 \pm 0.281 \leq 3.1190$ & $0.196 \pm 0.396 \leq 2.9770$ \\
$16$ & $0.090 \pm 0.253 \leq 2.9564$ & $0.168 \pm 0.366 \leq 2.5710$ \\
$17$ & $0.072 \pm 0.217 \leq 2.7127$ & $0.134 \pm 0.319 \leq 2.5699$ \\
$18$ & $0.052 \pm 0.175 \leq 2.3270$ & $0.093 \pm 0.252 \leq 2.1888$ \\
$19$ & $0.043 \pm 0.153 \leq 2.0983$ & $0.071 \pm 0.206 \leq 2.2149$ \\
$20$ & $0.027 \pm 0.103 \leq 0.8813$ & $0.050 \pm 0.167 \leq 2.1350$ \\
$21$ & $0.034 \pm 0.154 \leq 1.0307$ & $0.053 \pm 0.183 \leq 1.2459$ \\
$22$ & $0.000 \pm 0.000 \leq 0.0000$ & $0.000 \pm 0.000 \leq 0.0000$ \\
\hline
\end{tabular}

\caption{Locality statistics for the German fiction corpus}
\label{fictionlocality}
\end{table}

\skipAll{
\begin{table}
\center
\begin{tabular}{r|r|r|}
\ccnf{layer} & \cc{left} & \cc{right} \\
\hline
$1$ & $0.078 \pm 0.237 \leq 2.4615$ & $0.446 \pm 0.481 \leq 2.4615$ \\
$2$ & $0.151 \pm 0.321 \leq 2.9767$ & $0.560 \pm 0.515 \leq 2.7287$ \\
$3$ & $0.195 \pm 0.380 \leq 4.3580$ & $0.579 \pm 0.537 \leq 2.8638$ \\
$4$ & $0.212 \pm 0.386 \leq 3.4308$ & $0.552 \pm 0.542 \leq 2.9318$ \\
$5$ & $0.211 \pm 0.376 \leq 3.6215$ & $0.507 \pm 0.538 \leq 3.0283$ \\
$6$ & $0.196 \pm 0.364 \leq 3.2953$ & $0.453 \pm 0.528 \leq 3.0298$ \\
$7$ & $0.181 \pm 0.354 \leq 3.5694$ & $0.407 \pm 0.519 \leq 3.0461$ \\
$8$ & $0.163 \pm 0.336 \leq 3.3824$ & $0.356 \pm 0.500 \leq 2.9567$ \\
$9$ & $0.142 \pm 0.318 \leq 3.4881$ & $0.315 \pm 0.476 \leq 2.7772$ \\
$10$ & $0.126 \pm 0.297 \leq 3.2772$ & $0.274 \pm 0.451 \leq 2.7655$ \\
$11$ & $0.115 \pm 0.285 \leq 2.7124$ & $0.234 \pm 0.419 \leq 2.8320$ \\
$12$ & $0.094 \pm 0.258 \leq 2.7256$ & $0.205 \pm 0.406 \leq 2.7351$ \\
$13$ & $0.079 \pm 0.224 \leq 2.0900$ & $0.168 \pm 0.358 \leq 2.6583$ \\
$14$ & $0.070 \pm 0.207 \leq 2.1934$ & $0.152 \pm 0.339 \leq 2.3566$ \\
$15$ & $0.049 \pm 0.165 \leq 2.5056$ & $0.112 \pm 0.274 \leq 2.4065$ \\
$16$ & $0.031 \pm 0.123 \leq 2.7183$ & $0.072 \pm 0.202 \leq 1.8561$ \\
$17$ & $0.019 \pm 0.086 \leq 1.1320$ & $0.040 \pm 0.142 \leq 1.3769$ \\
$18$ & $0.026 \pm 0.113 \leq 0.9203$ & $0.083 \pm 0.207 \leq 0.9232$ \\
$19$ & $0.068 \pm 0.230 \leq 1.2137$ & $0.104 \pm 0.221 \leq 0.9577$ \\
$20$ & $0.045 \pm 0.110 \leq 0.3913$ & $0.147 \pm 0.260 \leq 0.9310$ \\
$21$ & $0.000 \pm 0.000 \leq 0.0000$ & $0.142 \pm 0.232 \leq 0.6939$ \\
\hline
\end{tabular} 

\caption{Locality statistics for the Java source code corpus}
\label{javalocality}
\end{table}
}

\begin{table}
\center
\begin{tabular}{r|r|r|}
\ccnf{layer} & \cc{left} & \cc{right} \\
\hline
$1$ & $0.096 \pm 0.270 \leq 2.9538$ & $0.452 \pm 0.485 \leq 1.9692$ \\
$2$ & $0.171 \pm 0.349 \leq 3.2248$ & $0.598 \pm 0.517 \leq 2.7287$ \\
$3$ & $0.206 \pm 0.389 \leq 3.4864$ & $0.625 \pm 0.540 \leq 2.8638$ \\
$4$ & $0.218 \pm 0.386 \leq 3.6179$ & $0.593 \pm 0.546 \leq 3.0565$ \\
$5$ & $0.216 \pm 0.379 \leq 3.1532$ & $0.541 \pm 0.542 \leq 2.9659$ \\
$6$ & $0.209 \pm 0.378 \leq 3.4671$ & $0.491 \pm 0.539 \leq 3.1547$ \\
$7$ & $0.195 \pm 0.366 \leq 3.9365$ & $0.446 \pm 0.532 \leq 3.2180$ \\
$8$ & $0.182 \pm 0.357 \leq 4.3588$ & $0.405 \pm 0.523 \leq 2.8825$ \\
$9$ & $0.167 \pm 0.344 \leq 3.9138$ & $0.367 \pm 0.508 \leq 2.8904$ \\
$10$ & $0.151 \pm 0.323 \leq 3.5809$ & $0.316 \pm 0.474 \leq 2.8271$ \\
$11$ & $0.124 \pm 0.292 \leq 2.9052$ & $0.252 \pm 0.430 \leq 2.8725$ \\
$12$ & $0.080 \pm 0.217 \leq 1.9145$ & $0.166 \pm 0.334 \leq 2.0134$ \\
$13$ & $0.031 \pm 0.104 \leq 0.8367$ & $0.060 \pm 0.150 \leq 0.8757$ \\
\hline
\end{tabular} 

\caption{Locality statistics for the random corpus}
\label{randomlocality}
\end{table}


\begin{table}
\center
\begin{tabular}{r|r|r|r|r|r|r|r|r|r|}
\ccnf{layer} & \multicolumn{2}{|c|}{Balancing phase} & \cc{Caterpillar} & \multicolumn{6}{|c|}{Diffbit phase} \\
 & \cc{$0$} & \cc{$1$} & \cc{phase} & \cc{$0$} & \cc{$1$} & \cc{$2$} & \cc{$3$} & \cc{$4$} & \cc{$5$}  \\
\hline
All  & $74.8294$  & $3.0070$  & $3.1737$  & $14.7623$  & $3.9450$  & $0.1886$  & $0.0180$  & $0.0732$  & $0.0028$ \\
\hline $1$  & $72.3374$  & $0.0085$  & $7.2716$  & $16.2139$  & $3.7982$  & $0.2343$  & $0.0251$  & $0.1097$  & $0.0012$ \\
 $2$  & $81.9343$  & $1.2419$  & $0.0805$  & $13.9402$  & $2.6007$  & $0.1608$  & $0.0006$  & $0.0410$  & $0.0001$ \\
 $3$  & $74.6017$  & $6.6423$  & $0.0430$  & $13.4103$  & $5.0654$  & $0.1639$  & $0.0161$  & $0.0521$  & $0.0053$ \\
 $4$  & $71.8794$  & $8.9373$  & $0.0156$  & $13.6783$  & $5.2664$  & $0.1439$  & $0.0232$  & $0.0490$  & $0.0070$ \\
 $5$  & $70.8688$  & $10.6544$  & $0.0093$  & $13.0719$  & $5.1969$  & $0.1224$  & $0.0266$  & $0.0412$  & $0.0085$ \\
 $6$  & $70.5503$  & $11.0393$  & $0.0078$  & $12.9915$  & $5.2102$  & $0.1234$  & $0.0283$  & $0.0406$  & $0.0086$ \\
 $7$  & $71.0367$  & $10.3319$  & $0.0058$  & $13.4759$  & $4.9324$  & $0.1306$  & $0.0321$  & $0.0445$  & $0.0101$ \\
 $8$  & $72.0536$  & $8.0441$  & $0.0006$  & $14.5473$  & $5.0916$  & $0.1598$  & $0.0387$  & $0.0495$  & $0.0148$ \\
 $9$  & $70.7695$  & $9.1087$  & $0.0004$  & $14.6739$  & $5.2104$  & $0.1413$  & $0.0411$  & $0.0432$  & $0.0116$ \\
 $10$  & $70.5434$  & $9.8697$  & $0.0002$  & $14.1661$  & $5.1930$  & $0.1382$  & $0.0352$  & $0.0428$  & $0.0116$ \\
 $11$  & $69.2616$  & $11.3420$  & $0.0000$  & $13.9973$  & $5.1663$  & $0.1368$  & $0.0366$  & $0.0462$  & $0.0132$ \\
 $12$  & $68.9641$  & $12.5867$  & $0.0000$  & $13.1362$  & $5.1078$  & $0.1182$  & $0.0337$  & $0.0429$  & $0.0104$ \\
 $13$  & $68.1771$  & $14.6208$  & $0.0000$  & $12.2444$  & $4.7717$  & $0.1096$  & $0.0298$  & $0.0371$  & $0.0095$ \\
 $14$  & $66.8986$  & $17.2181$  & $0.0000$  & $11.1095$  & $4.5963$  & $0.1054$  & $0.0333$  & $0.0279$  & $0.0108$ \\
 $15$  & $65.5028$  & $19.9586$  & $0.0000$  & $10.0405$  & $4.3778$  & $0.0711$  & $0.0197$  & $0.0257$  & $0.0039$ \\
 $16$  & $65.1442$  & $21.2371$  & $0.0000$  & $9.4443$  & $4.0218$  & $0.0925$  & $0.0139$  & $0.0324$  & $0.0139$ \\
 $17$  & $65.7033$  & $20.4522$  & $0.0000$  & $10.1853$  & $3.5660$  & $0.0350$  & $0.0350$  & $0.0117$  & $0.0117$ \\
 $18$  & $67.8603$  & $17.2486$  & $0.0000$  & $10.6535$  & $4.1779$  & $0.0597$  & $0.0000$  & $0.0000$  & $0.0000$ \\
 $19$  & $66.7128$  & $16.8858$  & $0.0000$  & $11.9723$  & $4.3599$  & $0.0000$  & $0.0000$  & $0.0000$  & $0.0692$ \\
 $20$  & $65.9021$  & $18.3486$  & $0.0000$  & $10.3976$  & $5.3517$  & $0.0000$  & $0.0000$  & $0.0000$  & $0.0000$ \\
 $21$  & $63.3898$  & $23.7288$  & $0.0000$  & $9.4915$  & $3.0508$  & $0.0000$  & $0.3390$  & $0.0000$  & $0.0000$ \\
 $22$  & $65.8537$  & $21.1382$  & $0.0000$  & $9.7561$  & $3.2520$  & $0.0000$  & $0.0000$  & $0.0000$  & $0.0000$ \\
 $23$  & $57.8947$  & $35.0877$  & $0.0000$  & $3.5088$  & $3.5088$  & $0.0000$  & $0.0000$  & $0.0000$  & $0.0000$ \\
 $24$  & $50.0000$  & $43.7500$  & $0.0000$  & $6.2500$  & $0.0000$  & $0.0000$  & $0.0000$  & $0.0000$  & $0.0000$ \\
\hline
\end{tabular}

\caption{Phase census for the kernel source code corpus}
\label{kernelcensus}
\end{table}

\begin{table}
\center
\begin{tabular}{r|r|r|r|r|r|r|r|r|r|}
\ccnf{layer} & \multicolumn{2}{|c|}{Balancing phase} & \cc{Caterpillar} & \multicolumn{6}{|c|}{Diffbit phase} \\
 & \cc{$0$} & \cc{$1$} & \cc{phase} & \cc{$0$} & \cc{$1$} & \cc{$2$} & \cc{$3$} & \cc{$4$} & \cc{$5$}  \\
\hline
All  & $74.9836$  & $2.1322$  & $1.8579$  & $16.6059$  & $4.1589$  & $0.1897$  & $0.0093$  & $0.0596$  & $0.0029$ \\
\hline $1$  & $70.3134$  & $0.0000$  & $4.2760$  & $20.0963$  & $4.9822$  & $0.2449$  & $0.0010$  & $0.0854$  & $0.0007$ \\
 $2$  & $85.5433$  & $0.0530$  & $0.0206$  & $13.0597$  & $1.1971$  & $0.1074$  & $0.0001$  & $0.0187$  & $0.0000$ \\
 $3$  & $75.1115$  & $4.8630$  & $0.0017$  & $14.4251$  & $5.3066$  & $0.2036$  & $0.0217$  & $0.0612$  & $0.0057$ \\
 $4$  & $72.0470$  & $7.7268$  & $0.0011$  & $14.7095$  & $5.2639$  & $0.1616$  & $0.0290$  & $0.0527$  & $0.0084$ \\
 $5$  & $71.5391$  & $8.1232$  & $0.0001$  & $14.7604$  & $5.3316$  & $0.1499$  & $0.0338$  & $0.0511$  & $0.0109$ \\
 $6$  & $70.9514$  & $8.6365$  & $0.0000$  & $14.8445$  & $5.3194$  & $0.1483$  & $0.0373$  & $0.0503$  & $0.0122$ \\
 $7$  & $70.6358$  & $8.8872$  & $0.0001$  & $14.9030$  & $5.3253$  & $0.1467$  & $0.0390$  & $0.0497$  & $0.0132$ \\
 $8$  & $70.4886$  & $8.9909$  & $0.0000$  & $14.9433$  & $5.3269$  & $0.1465$  & $0.0411$  & $0.0501$  & $0.0124$ \\
 $9$  & $70.4329$  & $9.0797$  & $0.0000$  & $14.9401$  & $5.2981$  & $0.1457$  & $0.0416$  & $0.0500$  & $0.0119$ \\
 $10$  & $70.3646$  & $9.1681$  & $0.0000$  & $14.9111$  & $5.3071$  & $0.1459$  & $0.0396$  & $0.0498$  & $0.0137$ \\
 $11$  & $70.3841$  & $9.2078$  & $0.0000$  & $14.8692$  & $5.2864$  & $0.1513$  & $0.0425$  & $0.0467$  & $0.0119$ \\
 $12$  & $70.2532$  & $9.4323$  & $0.0000$  & $14.7734$  & $5.2931$  & $0.1490$  & $0.0382$  & $0.0501$  & $0.0106$ \\
 $13$  & $70.0808$  & $9.7148$  & $0.0000$  & $14.7517$  & $5.2193$  & $0.1379$  & $0.0345$  & $0.0456$  & $0.0154$ \\
 $14$  & $69.8633$  & $10.4019$  & $0.0000$  & $14.2467$  & $5.2673$  & $0.1221$  & $0.0425$  & $0.0446$  & $0.0117$ \\
 $15$  & $69.4125$  & $11.4312$  & $0.0000$  & $13.7447$  & $5.1415$  & $0.1531$  & $0.0489$  & $0.0553$  & $0.0128$ \\
 $16$  & $68.6791$  & $13.2387$  & $0.0000$  & $12.8891$  & $4.9714$  & $0.1322$  & $0.0341$  & $0.0426$  & $0.0128$ \\
 $17$  & $67.7433$  & $16.7717$  & $0.0000$  & $10.6486$  & $4.7241$  & $0.0432$  & $0.0345$  & $0.0173$  & $0.0173$ \\
 $18$  & $64.6302$  & $22.3651$  & $0.0000$  & $8.8782$  & $4.0550$  & $0.0715$  & $0.0000$  & $0.0000$  & $0.0000$ \\
 $19$  & $59.0720$  & $32.7801$  & $0.0000$  & $5.0170$  & $3.0932$  & $0.0377$  & $0.0000$  & $0.0000$  & $0.0000$ \\
 $20$  & $59.4037$  & $35.8945$  & $0.0000$  & $2.6376$  & $2.0642$  & $0.0000$  & $0.0000$  & $0.0000$  & $0.0000$ \\
 $21$  & $48.5549$  & $49.1329$  & $0.0000$  & $1.1561$  & $1.1561$  & $0.0000$  & $0.0000$  & $0.0000$  & $0.0000$ \\
 $22$  & $68.7500$  & $25.0000$  & $0.0000$  & $6.2500$  & $0.0000$  & $0.0000$  & $0.0000$  & $0.0000$  & $0.0000$ \\
\hline
\end{tabular}

\caption{Phase census for the German fiction corpus}
\label{fictioncensus}
\end{table}

\skipAll{
\begin{table}
\center
\begin{tabular}{r|r|r|r|r|r|r|r|r|r|}
\ccnf{layer} & \multicolumn{2}{|c|}{Balancing phase} & \cc{Caterpillar} & \multicolumn{6}{|c|}{Diffbit phase} \\
 & \cc{$0$} & \cc{$1$} & \cc{phase} & \cc{$0$} & \cc{$1$} & \cc{$2$} & \cc{$3$} & \cc{$4$} & \cc{$5$}  \\
\hline
All  & $74.4794$  & $2.6480$  & $4.3800$  & $14.4232$  & $3.7970$  & $0.2080$  & $0.0086$  & $0.0534$  & $0.0024$ \\
\hline $1$  & $71.3905$  & $0.0037$  & $10.0668$  & $14.4547$  & $3.7787$  & $0.2396$  & $0.0013$  & $0.0642$  & $0.0006$ \\
 $2$  & $82.0168$  & $1.0072$  & $0.0373$  & $14.7837$  & $1.9037$  & $0.2217$  & $0.0009$  & $0.0286$  & $0.0000$ \\
 $3$  & $74.4891$  & $6.5271$  & $0.0857$  & $13.5832$  & $5.0699$  & $0.1598$  & $0.0149$  & $0.0669$  & $0.0033$ \\
 $4$  & $72.5579$  & $7.3615$  & $0.0330$  & $14.5453$  & $5.2558$  & $0.1601$  & $0.0291$  & $0.0497$  & $0.0077$ \\
 $5$  & $71.7215$  & $8.0887$  & $0.0080$  & $14.5530$  & $5.3784$  & $0.1561$  & $0.0308$  & $0.0541$  & $0.0094$ \\
 $6$  & $70.9952$  & $8.6595$  & $0.0061$  & $14.8575$  & $5.2539$  & $0.1356$  & $0.0339$  & $0.0468$  & $0.0116$ \\
 $7$  & $70.9705$  & $8.7717$  & $0.0012$  & $14.7521$  & $5.2794$  & $0.1479$  & $0.0339$  & $0.0313$  & $0.0121$ \\
 $8$  & $70.2738$  & $9.4320$  & $0.0006$  & $14.7587$  & $5.2860$  & $0.1559$  & $0.0350$  & $0.0449$  & $0.0131$ \\
 $9$  & $70.1256$  & $9.7127$  & $0.0000$  & $14.6469$  & $5.2998$  & $0.1162$  & $0.0530$  & $0.0411$  & $0.0047$ \\
 $10$  & $70.4221$  & $9.7976$  & $0.0000$  & $14.5313$  & $5.0055$  & $0.1333$  & $0.0636$  & $0.0399$  & $0.0066$ \\
 $11$  & $69.2438$  & $10.8260$  & $0.0000$  & $14.6368$  & $5.0861$  & $0.1133$  & $0.0473$  & $0.0372$  & $0.0095$ \\
 $12$  & $68.4196$  & $12.1855$  & $0.0000$  & $13.9145$  & $5.1338$  & $0.2283$  & $0.0436$  & $0.0742$  & $0.0005$ \\
 $13$  & $68.9905$  & $14.0328$  & $0.0000$  & $11.3125$  & $5.5487$  & $0.1081$  & $0.0042$  & $0.0021$  & $0.0011$ \\
 $14$  & $67.6668$  & $17.0817$  & $0.0000$  & $9.9167$  & $5.2180$  & $0.0151$  & $0.0800$  & $0.0043$  & $0.0173$ \\
 $15$  & $63.0015$  & $23.7091$  & $0.0000$  & $7.2555$  & $5.9899$  & $0.0308$  & $0.0132$  & $0.0000$  & $0.0000$ \\
 $16$  & $59.7072$  & $30.7467$  & $0.0000$  & $6.7350$  & $2.8111$  & $0.0000$  & $0.0000$  & $0.0000$  & $0.0000$ \\
 $17$  & $52.1282$  & $43.0625$  & $0.0000$  & $0.8015$  & $4.0077$  & $0.0000$  & $0.0000$  & $0.0000$  & $0.0000$ \\
 $18$  & $50.1779$  & $47.0937$  & $0.0000$  & $2.2539$  & $0.4745$  & $0.0000$  & $0.0000$  & $0.0000$  & $0.0000$ \\
 $19$  & $51.3514$  & $37.8378$  & $0.0000$  & $8.1081$  & $2.7027$  & $0.0000$  & $0.0000$  & $0.0000$  & $0.0000$ \\
 $20$  & $55.5556$  & $44.4444$  & $0.0000$  & $0.0000$  & $0.0000$  & $0.0000$  & $0.0000$  & $0.0000$  & $0.0000$ \\
 $21$  & $50.0000$  & $50.0000$  & $0.0000$  & $0.0000$  & $0.0000$  & $0.0000$  & $0.0000$  & $0.0000$  & $0.0000$ \\
 $22$  & $0.0000$  & $100.0000$  & $0.0000$  & $0.0000$  & $0.0000$  & $0.0000$  & $0.0000$  & $0.0000$  & $0.0000$ \\
\hline
\end{tabular}

\caption{Phase census for the Java source code corpus}
\label{javacensus}
\end{table}
}

\begin{table}
\center
\begin{tabular}{r|r|r|r|r|r|r|r|r|r|}
\ccnf{layer} & \multicolumn{2}{|c|}{Balancing phase} & \cc{Caterpillar} & \multicolumn{6}{|c|}{Diffbit phase} \\
 & \cc{$0$} & \cc{$1$} & \cc{phase} & \cc{$0$} & \cc{$1$} & \cc{$2$} & \cc{$3$} & \cc{$4$} & \cc{$5$}  \\
\hline
All  & $76.8752$  & $2.2763$  & $0.2617$  & $16.4472$  & $3.7615$  & $0.2693$  & $0.0139$  & $0.0906$  & $0.0043$ \\
\hline $1$  & $76.7821$  & $0.0067$  & $0.6062$  & $18.4574$  & $3.6395$  & $0.3481$  & $0.0116$  & $0.1448$  & $0.0036$ \\
 $2$  & $82.0945$  & $0.0176$  & $0.0001$  & $15.5173$  & $2.0647$  & $0.2630$  & $0.0017$  & $0.0407$  & $0.0005$ \\
 $3$  & $74.5109$  & $5.8501$  & $0.0000$  & $14.2281$  & $5.1421$  & $0.1825$  & $0.0206$  & $0.0596$  & $0.0060$ \\
 $4$  & $72.3018$  & $7.3711$  & $0.0000$  & $14.7577$  & $5.3116$  & $0.1644$  & $0.0278$  & $0.0564$  & $0.0092$ \\
 $5$  & $71.5582$  & $8.1825$  & $0.0000$  & $14.6944$  & $5.3168$  & $0.1504$  & $0.0331$  & $0.0540$  & $0.0105$ \\
 $6$  & $70.8267$  & $8.8355$  & $0.0000$  & $14.7880$  & $5.3010$  & $0.1506$  & $0.0376$  & $0.0487$  & $0.0118$ \\
 $7$  & $70.5169$  & $9.1850$  & $0.0000$  & $14.7491$  & $5.3070$  & $0.1463$  & $0.0369$  & $0.0473$  & $0.0116$ \\
 $8$  & $70.2640$  & $9.6263$  & $0.0000$  & $14.5724$  & $5.2913$  & $0.1471$  & $0.0385$  & $0.0479$  & $0.0125$ \\
 $9$  & $69.9518$  & $10.2376$  & $0.0000$  & $14.3149$  & $5.2571$  & $0.1392$  & $0.0383$  & $0.0490$  & $0.0122$ \\
 $10$  & $69.4257$  & $11.6426$  & $0.0000$  & $13.4317$  & $5.2945$  & $0.1191$  & $0.0421$  & $0.0364$  & $0.0078$ \\
 $11$  & $68.1849$  & $14.2384$  & $0.0000$  & $12.3194$  & $5.0931$  & $0.1042$  & $0.0186$  & $0.0343$  & $0.0071$ \\
 $12$  & $66.0665$  & $19.5077$  & $0.0000$  & $9.3351$  & $4.9994$  & $0.0541$  & $0.0028$  & $0.0342$  & $0.0000$ \\
 $13$  & $71.7154$  & $26.9588$  & $0.0000$  & $1.3258$  & $0.0000$  & $0.0000$  & $0.0000$  & $0.0000$  & $0.0000$ \\
 $14$  & $58.9300$  & $41.0700$  & $0.0000$  & $0.0000$  & $0.0000$  & $0.0000$  & $0.0000$  & $0.0000$  & $0.0000$ \\
\hline
\end{tabular} 

\caption{Phase census for the random corpus}
\label{randomcensus}
\end{table}

\skipAll{
\begin{table}
\center
\begin{tabular}{r|r|r|r|r|r|}
layer & \#chunks & \#unique & caterpillars & compressed & ratio \\
\hline
$0$  & $1.34\cdot 10^{9}$  & $378$ & $0$ & $1.43\cdot 10^{9}$ & $1.063$\\
$1$  & $6.50\cdot 10^{8}$  & $11699$ & $793$ & $1.10\cdot 10^{9}$ & $0.813$\\
$2$  & $3.76\cdot 10^{8}$  & $873439$ & $1330$ & $9.31\cdot 10^{8}$ & $0.686$\\
$3$  & $2.02\cdot 10^{8}$  & $1.10\cdot 10^{7}$ & $2225$ & $6.69\cdot 10^{8}$ & $0.492$\\
$4$  & $1.06\cdot 10^{8}$  & $2.85\cdot 10^{7}$ & $2319$ & $6.86\cdot 10^{8}$ & $0.504$\\
$5$  & $5.63\cdot 10^{7}$  & $2.84\cdot 10^{7}$ & $1752$ & $8.37\cdot 10^{8}$ & $0.615$\\
$6$  & $2.97\cdot 10^{7}$  & $1.87\cdot 10^{7}$ & $1096$ & $9.43\cdot 10^{8}$ & $0.693$\\
$7$  & $1.48\cdot 10^{7}$  & $1.09\cdot 10^{7}$ & $670$ & $1.04\cdot 10^{9}$ & $0.764$\\
$8$  & $7.54\cdot 10^{6}$  & $5.82\cdot 10^{6}$ & $339$ & $1.07\cdot 10^{9}$ & $0.783$\\
$9$  & $3.81\cdot 10^{6}$  & $3.00\cdot 10^{6}$ & $181$ & $1.08\cdot 10^{9}$ & $0.795$\\
$10$  & $1.92\cdot 10^{6}$  & $1.55\cdot 10^{6}$ & $86$ & $1.10\cdot 10^{9}$ & $0.810$\\
$11$  & $962532$  & $797752$ & $41$ & $1.13\cdot 10^{9}$ & $0.829$\\
$12$  & $491352$  & $420302$ & $16$ & $1.16\cdot 10^{9}$ & $0.852$\\
$13$  & $259613$  & $230635$ & $4$ & $1.20\cdot 10^{9}$ & $0.879$\\
$14$  & $148654$  & $137435$ & $0$ & $1.24\cdot 10^{9}$ & $0.908$\\
$15$  & $97989$  & $93958$ & $0$ & $1.27\cdot 10^{9}$ & $0.935$\\
$16$  & $76357$  & $74910$ & $0$ & $1.30\cdot 10^{9}$ & $0.957$\\
$17$  & $67776$  & $67218$ & $0$ & $1.32\cdot 10^{9}$ & $0.971$\\
$18$  & $64425$  & $64159$ & $0$ & $1.33\cdot 10^{9}$ & $0.980$\\
$19$  & $62980$  & $62850$ & $0$ & $1.34\cdot 10^{9}$ & $0.988$\\
$20$  & $62326$ & $62265$ & $0$ & $1.36\cdot 10^{9}$ & $0.996$\\
$21$  & $62031$ & $61976$ & $0$ & $1.36\cdot 10^{9}$ & $0.998$\\
$22$  & $61908$ & $61865$ & $0$ & $1.36\cdot 10^{9}$ & $0.998$\\
$23$  & $61851$ & $61812$ & $0$ & $1.36\cdot 10^{9}$ & $1.000$\\
$24$  & $61835$ & $61796$ & $0$ & $1.36\cdot 10^{9}$ & $1.000$\\
$25$  & $61834$ & $61795$ & $0$ & $1.36\cdot 10^{9}$ & $1.000$\\
\hline
\end{tabular} 

\caption{Deduplication performance on the kernel source code corpus. The total UTF-8-encoded length of the data is $1\;347\;776\;301$ bytes.}
\label{kerneldedup}
\end{table}

\begin{table}
\center
\begin{tabular}{r|r|r|r|r|r|}
layer & \#chunks & \#unique & caterpillars & compressed & ratio \\
\hline
$0$  & $1.06\cdot 10^{9}$ & $353$ & $0$ & $1.12\cdot 10^{9}$ & $1.023$\\
$1$  & $5.97\cdot 10^{8}$ & $7975$ & $233$ & $9.67\cdot 10^{8}$ & $0.892$\\
$2$  & $3.44\cdot 10^{8}$  & $246991$ & $357$ & $7.71\cdot 10^{8}$ & $0.704$\\
$3$  & $1.82\cdot 10^{8}$  & $7.67\cdot 10^{6}$ & $506$ & $5.80\cdot 10^{8}$ & $0.527$\\
$4$  & $9.34\cdot 10^{7}$  & $4.62\cdot 10^{7}$ & $452$ & $9.17\cdot 10^{8}$ & $0.832$\\
$5$  & $4.75\cdot 10^{7}$  & $4.51\cdot 10^{7}$ & $204$ & $1.22\cdot 10^{9}$ & $1.103$\\
$6$  & $2.39\cdot 10^{7}$  & $2.36\cdot 10^{7}$ & $77$ & $1.16\cdot 10^{9}$ & $1.055$\\
$7$  & $1.20\cdot 10^{7}$  & $1.19\cdot 10^{7}$ & $28$ & $1.13\cdot 10^{9}$ & $1.023$\\
$8$  & $6.02\cdot 10^{6}$  & $5.98\cdot 10^{6}$ & $1$ & $1.11\cdot 10^{9}$ & $1.010$\\
$9$  & $3.01\cdot 10^{6}$  & $3.00\cdot 10^{6}$ & $0$ & $1.11\cdot 10^{9}$ & $1.004$\\
$10$  & $1.51\cdot 10^{6}$  & $1.51\cdot 10^{6}$ & $0$ & $1.10\cdot 10^{9}$ & $1.002$\\
$11$  & $754515$  & $753773$ & $0$ & $1.10\cdot 10^{9}$ & $1.001$\\
$12$  & $377365$  & $377143$ & $0$ & $1.10\cdot 10^{9}$ & $1.000$\\
$13$  & $188797$  & $188734$ & $0$ & $1.10\cdot 10^{9}$ & $1.000$\\
$14$  & $94593$  & $94581$ & $0$ & $1.10\cdot 10^{9}$ & $1.000$\\
$15$  & $47564$  & $47562$ & $0$ & $1.10\cdot 10^{9}$ & $1.000$\\
$16$  & $24110$  & $24110$ & $0$ & $1.10\cdot 10^{9}$ & $1.000$\\
$17$  & $12531$  & $12531$ & $0$ & $1.10\cdot 10^{9}$ & $1.000$\\
$18$  & $6933$  & $6933$ & $0$ & $1.10\cdot 10^{9}$ & $1.000$\\
$19$  & $4282$  & $4282$ & $0$ & $1.10\cdot 10^{9}$ & $1.000$\\
$20$  & $3410$  & $3410$ & $0$ & $1.10\cdot 10^{9}$ & $1.000$\\
$21$  & $3237$  & $3237$ & $0$ & $1.10\cdot 10^{9}$ & $1.000$\\
$22$  & $3221$  & $3221$ & $0$ & $1.10\cdot 10^{9}$ & $1.000$\\
$23$  & $3219$  & $3219$ & $0$ & $1.10\cdot 10^{9}$ & $1.000$\\
\hline
\end{tabular}

\caption{Deduplication performance on the German fiction corpus. The total UTF-8-encoded length of the data is $1\;093\;335\;012$ bytes.}
\label{fictiondedup}
\end{table}
}

\begin{table}
\center
\begin{tabular}{l|r|r|l|r|r|
}
Method \;& \; Size after dedup \;&\; Avg. chunk size \;&\; Dedup ratio \;&\; \#chunks \;&\; $\sigma$ chunk size
\\
\hline
AE-Max       & $2\,491\,689\,742$ &  $8\,639$ & $4.348$ & $1\,254\,056$ & $2\,525$ \\
AE-Min       & $2\,708\,751\,657$ &  $7\,890$ & $4$     & $1\,373\,112$ & $1\,659$ \\
(Chonkers 8k)& $2\,068\,662\,870$ &  $5\,663$ & $5.237$ & $1\,913\,012$ & $1\,479$ \\ 
Chonkers 12k & $2\,298\,410\,180$ &  $8\,475$ & $4.714$ & $1\,278\,218$ & $2\,227$ \\
CRC32        & $2\,160\,021\,602$ &  $8\,158$ & $5.016$ & $1\,327\,938$ & $6\,883$ \\
FastCDC      & $2\,190\,794\,217$ & $10\,402$ & $4.945$ & $1\,041\,496$ & $5\,374$ \\
Gear         & $2\,203\,885\,698$ &  $9\,608$ & $4.916$ & $1\,127\,564$ & $7\,652$ \\
MAXP         & $2\,043\,705\,107$ &  $6\,596$ & $5.301$ & $1\,642\,488$ & $8\,719$ \\
Rabin        & $2\,229\,576\,125$ & $10\,343$ & $4.859$ & $1\,047\,411$ & $8\,410$ \\
RAM          & $2\,514\,289\,924$ &  $9\,845$ & $4.309$ & $1\,100\,411$ & $5\,662$ \\
TTTD         & $2\,162\,210\,908$ &  $9\,682$ & $5.011$ & $1\,118\,924$ & $7\,784$ \\
\hline
\end{tabular}

\caption{Deduplication comparison. See \cite{liu2023dedupbench,github_dedup-bench_2025} for the other chunking methods. Sizes are in given in bytes. The un-deduplicated data size in all cases is $10\,834\,129\,277$ bytes}
\label{dedupBench}
\end{table}

\FloatBarrier

\end{document}